\documentclass[twocolumn]{aastex63}

\usepackage{amsmath}
\usepackage{amssymb}
\usepackage{euscript}
\usepackage{algorithm2e,algorithmic}
\usepackage{bbold}
\usepackage{paralist}
\usepackage{color}
\usepackage{amsmath}
\usepackage{lineno}
\usepackage{lineno}

\usepackage{comment}
\submitjournal{ApJ}

\shorttitle{A UV survey of ZTF SNe IIn}
\shortauthors{Soumagnac et al.}

\begin{document}

%\title{Supernova shock breakout through an aspherical wind:\
%the case of 2018\,fif}

%\title{The first Ultra-Violet Survey of type II\lowercase{n} supernovae constrains the geometry of their surrounding circumstellar material}

\title{Early Ultra-Violet observations of type II\lowercase{n} supernovae constrain the asphericity of their circumstellar material}

%\title{Constrain on the asphericity of type II\lowercase{n} supernovae circumstellar material from early Ultra-Violet observations}

\correspondingauthor{Maayane T. Soumagnac}
\email{mtsoumagnac@lbl.gov}
\author[0000-0001-6753-1488]{Maayane T. Soumagnac}
\affiliation{Lawrence Berkeley National Laboratory, 1 Cyclotron Road, Berkeley, CA 94720, USA}
\affiliation{Department of Particle Physics and Astrophysics, Weizmann Institute of Science, Rehovot 76100, Israel}
\author{Eran O. Ofek}
\affiliation{Department of Particle Physics and Astrophysics, Weizmann Institute of Science, Rehovot 76100, Israel}
\author{Jingyi Liang}
\affiliation{Department of Particle Physics and Astrophysics, Weizmann Institute of Science, Rehovot 76100, Israel}
\author{Avishay Gal-yam}
\affiliation{Department of Particle Physics and Astrophysics, Weizmann Institute of Science, Rehovot 76100, Israel}
\author{Peter Nugent}
\affiliation{Lawrence Berkeley National Laboratory, 1 Cyclotron Road, Berkeley, CA 94720, USA}
\affiliation{Department of Astronomy, University of California, Berkeley, CA 94720-3411, USA}
%\author{Eli Waxman}
%\affiliation{Department of Particle Physics and Astrophysics, Weizmann Institute of Science, Rehovot 76100, Israel}
\author{Yi Yang}
\affiliation{Department of Particle Physics and Astrophysics, Weizmann Institute of Science, Rehovot 76100, Israel}
\author[0000-0003-1673-970X]{S. Bradley Cenko}
\affiliation{Astrophysics Science Division, NASA Goddard Space Flight Center, MC 661, Greenbelt, MD 20771, USA}
\affiliation{Joint Space-Science Institute, University of Maryland, College Park, MD 20742, USA.}
\author{Jesper Sollerman}
\affiliation{The Oskar Klein Centre, Department of Astronomy, Stockholm University, AlbaNova, 10691 Stockholm, Sweden}
\author{Daniel A. Perley}
\affiliation{Astrophysics Research Institute, Liverpool John Moores University, 146 Brownlow Hill, Liverpool L3 5RF, UK}
\author{Igor Andreoni}
\affiliation{California Institute of Technology, 1200 East California Boulevard, MC 278-17, Pasadena, CA 91125, USA}
\author{Cristina Barbarino}
\affiliation{The Oskar Klein Centre, Department of Astronomy, Stockholm University, AlbaNova, 10691 Stockholm, Sweden}
\author{Kevin B. Burdge}
\affiliation{California Institute of Technology, 1200 East California Boulevard, MC 278-17, Pasadena, CA 91125, USA}
\author[0000-0001-8208-2473]{Rachel J. Bruch}
\affiliation{Department of Particle Physics and Astrophysics, Weizmann Institute of Science, Rehovot 76100, Israel}
\author[0000-0002-8989-0542]{Kishalay De}
\affiliation{California Institute of Technology, 1200 East California Boulevard, MC 278-17, Pasadena, CA 91125, USA}
\author{Alison Dugas}
\affiliation{California Institute of Technology, 1200 East California Boulevard, MC 278-17, Pasadena, CA 91125, USA}
\author{Christoffer Fremling}
\affiliation{California Institute of Technology, 1200 East California Boulevard, MC 278-17, Pasadena, CA 91125, USA}
\author[0000-0002-9154-3136]{Melissa L. Graham}
%\affiliation{Department of Astronomy, University of Washington, Box 351580, U.W., Seattle, WA 98195-1580, USA}
\affiliation{DIRAC Institute, Department of Astronomy, University of
Washington, Seattle, WA 98195, USA}
\author{Matthew J. Hankins}
\affiliation{California Institute of Technology, 1200 East California Boulevard, MC 278-17, Pasadena, CA 91125, USA}
\author{Nora Linn Strotjohann}
\affiliation{Department of Particle Physics and Astrophysics, Weizmann Institute of Science, Rehovot 76100, Israel}
\author{Shane Moran}
\affiliation{Tuorla Observatory, Department of Physics and Astronomy, FI-20014, University of Turku, Finland}
\affiliation{Nordic Optical Telescope, Apartado 474, E-38700 Santa Cruz de La Palma, Spain}
\author[0000-0002-0466-1119]{James D. Neill}
\affiliation{California Institute of Technology, 1200 East California Boulevard, MC 278-17, Pasadena, CA 91125, USA}
\author{Steve Schulze}
\affiliation{Department of Particle Physics and Astrophysics, Weizmann Institute of Science, Rehovot 76100, Israel}
\author[0000-0003-4401-0430]{David L. Shupe}
\affiliation{IPAC, California Institute of Technology, 1200 E. California Blvd, Pasadena, CA 91125, USA}
\author[0000-0002-3713-6337]{Brigitta M.\ Sip\H{o}cz}
\affiliation{DIRAC Institute, Department of Astronomy, University of Washington, Seattle, WA 98195, USA}
\author[0000-0002-5748-4558]{Kirsty Taggart}
\affiliation{Astrophysics Research Institute, Liverpool John Moores University, 146 Brownlow Hill, Liverpool L3 5RF, UK}
\author[0000-0003-3433-1492]{Leonardo Tartaglia}
\affiliation{The Oskar Klein Centre, Department of Astronomy, Stockholm University, AlbaNova, 10691 Stockholm, Sweden}
\author{Richard Walters}
\affiliation{California Institute of Technology, 1200 East California Boulevard, MC 278-17, Pasadena, CA 91125, USA}
\author{Lin Yan}
\affiliation{California Institute of Technology, 1200 East California Boulevard, MC 278-17, Pasadena, CA 91125, USA}
\author[0000-0001-8018-5348]{Yuhan Yao}
\affiliation{California Institute of Technology, 1200 East California Boulevard, MC 278-17, Pasadena, CA 91125, USA}
\author{Ofer Yaron}
\affiliation{Department of Particle Physics and Astrophysics, Weizmann Institute of Science, Rehovot 76100, Israel}
\author[0000-0001-8018-5348]{Eric C. Bellm}
\affiliation{DIRAC Institute, Department of Astronomy, University of Washington, 3910 15th Avenue NE, Seattle, WA 98195, USA} 
\author{Chris Cannella}
\affiliation{Department of Electrical and Computer Engineering, Duke University, Durham, North Carolina 27708, United States}
\author{Richard Dekany}
\affiliation{Caltech Optical Observatories, California Institute of Technology, Pasadena, CA}
\author[0000-0001-5060-8733]{Dmitry A. Duev}
\affiliation{Division of Physics, Mathematics and Astronomy, California Institute of Technology, Pasadena, CA 91125, USA}
\author{Michael Feeney}
\affiliation{Caltech Optical Observatories, California Institute of Technology, Pasadena, CA}
\author[0000-0001-9676-730X]{Sara Frederick}
\affiliation{Department of Astronomy, University of Maryland College Park, College Park, MD 20742, USA}
\author{Matthew J. Graham}
\affiliation{Division of Physics, Mathematics and Astronomy, California Institute of Technology, Pasadena, CA 91125, USA}
\author[0000-0003-2451-5482]{Russ R. Laher}
\affiliation{IPAC, California Institute of Technology, 1200 E. California Blvd, Pasadena, CA 91125, USA}
\author[0000-0002-8532-9395]{Frank J. Masci}
\affiliation{IPAC, California Institute of Technology, 1200 E. California Blvd, Pasadena, CA 91125, USA}
\author{Mansi M. Kasliwal}
\affiliation{Division of Physics, Mathematics and Astronomy, California Institute of Technology, Pasadena, CA 91125, USA}
\author[0000-0001-8594-8666]{Marek Kowalski}
\affiliation{DESY, 15738 Zeuthen, Germany
Institut für Physik, Humboldt-Universität zu Berlin, 12489 Berlin, Germany}
\author[0000-0001-9515-478X]{Adam A. Miller}
\affiliation{Center for Interdisciplinary Exploration and Research in Astrophysics (CIERA) and Department of Physics and Astronomy, Northwestern University, 2145 Sheridan Road, Evanston, IL 60208, USA}
\affiliation{The Adler Planetarium, Chicago, IL 60605, USA}
\author[0000-0002-8121-2560]{Mickael Rigault}
\affiliation{Universite Clermont Auvergne, CNRS/IN2P3, Laboratoire de Physique de Clermont, F-63000 Clermont-Ferrand, France.}
\author[0000-0001-7648-4142]{Ben Rusholme}
\affiliation{IPAC, California Institute of Technology, 1200 E. California Blvd, Pasadena, CA 91125, USA}

\begin{abstract}
\begin{comment}
We present a survey of the early evolution of Type IIn supernovae (SNe IIn) in the Ultra-Violet (UV) and visible light and show that at least one third of them appear to explode in aspherical circumstelllar clouds. Our sample consists of 12 SNe IIn discovered and observed with the Zwicky Transient Facility (ZTF) and followed-up in the UV with the {\it Neil Gehrels Swift Observatory}. We use these observations to constrain the geometry of the circumstellar material (CSM) surrounding SN IIn explosions, which may shed light on their progenitor diversity. Indeed, while observations of SNe IIn are usually analyzed within the framework of spherically symmetric models of CSM, resolved images of stars undergoing considerable mass loss suggest that asphericity is common, and should be taken into account for realistic modeling of these events. We apply the criterion for asphericity introduced by Soumagnac et al., stating that a fast increase of the blackbody effective radius, if observed at times when the CSM surrounding the explosion is still optically thick, may be interpreted as an indication that the CSM is aspherical. We find that two thirds of the SNe in our sample show evidence for aspherical CSM, whereas one third do not show such evidence. After correcting for the relative volume of these two sub-classes, we derive a lower limit of 35\% on the fraction of SNe IIn showing evidence for aspherical CSM. 
\end{comment}

We present a survey of the early evolution of 12 Type IIn supernovae (SNe IIn) in the Ultra-Violet (UV) and visible light. %Our sample consists of SNe IIn discovered and observed with the Zwicky Transient Facility (ZTF) and followed-up in the UV with the Neil Gehrels Swift Observatory. 
We use this survey to constrain the geometry of the circumstellar material (CSM) surrounding SN IIn explosions, which may shed light on their progenitor diversity. %Indeed, while observations of SNe IIn are usually analyzed within the framework of spherically symmetric models of CSM, resolved images of stars undergoing considerable mass loss suggest that asphericity is common, and should be taken into account for realistic modeling of these events. 
In order to distinguish between aspherical and spherical circumstellar material (CSM), we estimate the blackbody radius temporal evolution of the SNe IIn of our sample, following the method introduced by Soumagnac et al. We find that higher luminosity objects tend to show evidence for aspherical CSM. Depending on whether this correlation is due to physical reasons or to some selection bias, we derive a lower limit between $35\%$ and $66\%$ on the fraction of SNe IIn showing evidence for aspherical CSM. This result suggests that asphericity of the CSM surrounding SNe IIn is common -- consistent with data from resolved images of stars undergoing considerable mass loss. It should be taken into account for more realistic modelling of these events.

%a correlation which could be due to either physical reasons or to some selection bias. If this correlation is physical, at least two third of SNe IIn could be exploding in aspherical CSM. However, if it is due instead to selection effects, we show that the lower limit on the fraction of SNe IIn showing evidence for aspherical CSM is $35\%$. In both cases, our analysis suggests that asphericity of the CSM is common arround SNe IIn, confirming the data from resolved images of stars undergoing considerable mass loss, and that it should be taken into account for more realistic modeling of these events. 

\end{abstract}

%globular clusters: general --- globular clusters: individual(NGC 6397,
%NGC 6624, NGC 7078, Terzan 8}

\section{Introduction}

Type IIn supernovae (SNe IIn) show prominent and narrow-to-intermediate width Balmer emission lines in their spectra \citep{Schlegel1990, Filippenko1997, Smith2014,Gal-Yam2017}. This specificity is thought to be the signature of photoionized and dense, hydrogen-rich, circumstellar medium (CSM) which is ejected from the SN progenitor prior to its explosive death.  Because they are the signature of an external physical phenomenon rather than of any intrinsic property of the explosion, these narrow lines may appear in the spectra of many SNe, at some point during their evolution. As a result, the Type IIn class of SNe is a heterogeneous category of objects. Depending on the spatial distribution and physical properties of the CSM surrounding the explosion, the characteristic narrow Balmer lines may persist for days (``flash spectroscopy'', \citealt{Gal-Yam2014,Khazov2016,Yaron2017}), weeks (e.g., SN\,1998S, \citealt{Li1998, Fassia2000, Fassia2001}; SN\,2005gl, \citealt{Gal-Yam2007}; SN\,2010mc, \citealt{Ofek2013}), or years (e.g., SN\,1988Z, \citealt{Danziger1991,Stathakis1991,Turatto1993,VanDyk1993,Chugai1994,Fabian1996,Aretxaga1999,Williams2002,Schlegel2006,Smith2017}; SN\,2010jl, \citealt{Patat2011, Stoll2011, Gall2014,Ofek2014}).

Observing SNe IIn at ultraviolet (UV) wavelengths is interesting for several reasons. First, an important ingredient of the physical picture governing SNe IIn explosions - the collisionless shock propagating in the CSM after the shock breakout \citep{Ofek2010} -- is predicted to radiate most in the UV and X-rays \citep{Katz2011,Murase2011,Murase2014,Chevalier2012}. Observing the explosion at these wavelengths has the potential to unveil precious information about the explosion mechanism and the CSM properties (e.g., \citealt{Ofek2013a}). In particular, it may provide a much better estimate of the bolometric luminosity of the event. 

Second, UV observations can help constrain the geometrical distribution of the CSM, which is closely related to the mass-loss processes occurring before the explosion and probe the nature of the progenitors of this type of events.

Although observations of SNe IIn are usually analyzed within the framework of spherically symmetric models of CSM, resolved images of stars undergoing considerable mass loss (e.g., $\eta$ Carinae; \citealt{Davidson1997, Davidson2012}), some of whom are probably SN IIn progenitors \citep{Gal-Yam2007,Gal-Yam2009} as well as polarimetric observations \citep{Leonard2000, Hoffman2008, Wang2008, Reilly2017} suggest that asphericity should be taken into account for more realistic modeling. Asphericity of the CSM has recently been invoked to interpret the spectrocopic and spectropolarimetric observations of the Type IIn SN\,2012ab \citep{Bilinski2017} and SN\,2009ip \citep{Mauerhan2014,Smith2014_2009ip,Levesque2014,Reilly2017}. 

In \cite{Soumagnac2019}, we showed that the light curve of the luminous Type IIn SN PTF\,12glz may be interpreted as evidence for aspherical CSM. While the spectroscopic analysis is consistent with opaque CSM obstructing our view of any growing structure, $r_{BB}$ - the radius of the deepest transparent emitting layer -- grows by an order of magnitude, at a speed of $\sim 8000\, \rm km\,s^{-1}$. 
%In addition to being inconsistent with the spectroscopic observations, the rapidly growing blackbody radius $r_{BB}$ of PTF\,12glz is also in contradiction with  most previous observations of SNe IIn. Some of the SNe IIn observed in the UV by \cite{DelaRosa2016} showed radii growing at comparable rates. All other observed SNe IIn seem to exhibit either a constant blackbody radius (e.g., SN\,2010jl \citealt{Ofek2014}), a blackbody radius stalling after a short increase (e.g., SNe\,2005kj, 2006bo, 2008fq, 2006qq, \citealt{Taddia2013}; SN\,2006tf, \citealt{Smith2008}) or even a shrinking blackbody radius (e.g., SNe\,2005ip; 2006jd, \citealt{Taddia2013}). 
To explain this tension, we considered a simple aspherical structure of CSM: a three-dimensional slab, infinite in two dimensions ($x$ and $y$ axis) and perpendicular to the line of sight ($z$ axis). We modeled the radiation from an explosion embedded in a slab of CSM by numerically solving the radiative diffusion equation in a slab with different density profiles: $\rho=Const.$, $\rho\propto |z|^{-1}$ and a wind density profile $\rho\propto z^{-2}$. Although this model is simplistic, it allows recovery of the peculiar growth of the  blackbody radius $r_{BB}$ observed in the case of PTF\,12glz, as well as the decrease of its blackbody temperature $T_{BB}$. 

This allowed us to derive a criterion for asphericity: a fast increase of $r_{BB}$ can be interpreted as the signature of non-spherical CSM, if it is observed while the CSM is still optically thick. This is because the approximately stationary CSM is obscuring the expanding SN ejecta, and explaining an expanding emitting region due to photon diffusion in the CSM requires a non-spherical CSM configuration. In this paper, we assemble a sample of SNe IIn, to which we apply this criterion in order to estimate the fraction of SNe IIn showing evidence for non-spherical CSM.

Several samples of SNe IIn have been gathered and studied so far. Among them, the sample by \cite{Kiewe2012}, consists of four SNe IIn observed by the Caltech Core-Collapse Project (CCCP) with the 1.5 m robotic telescope at the Palomar Observatory (P60; Cenko et al. 2006) using Johnson-Cousins $BVRI$ filters. They studies the light curve features and derived the progenitor star wind velocities. The sample by \cite{Taddia2013} consists of five SNe IIn observed by the Carnegie Supernova Project \citep{Hamuy2006} at visible-light and near-infrared wavelengths, and was used to derive mass-loss parameters. The sample by \cite{Ofek2014} consists of 19 SNe IIn observed by the Palomar Transient Factory (Law et al. 2009; Rau et al. 2009) and its extension, the intermediate PTF (iPTF) using the PTF $R$-band filter. It allowed to exhibit a possible correlation between the $r$-band rise time and peak luminosity of SNe IIn and to derive a lower limits on the shock-breakout velocity, supporting the idea that early-time light curves of SNe IIn are caused by shock breakout in a dense CSM. The sample by \cite{Nyholm2019} consists of $42$ objects with observations from PTF and iPTF, and was used for an in-depth study of their light-curve properties. \cite{DelaRosa2016} collected Swift UV observations of ten SNe IIn observed between 2007 and 2013 (eight of which post-peak) and studied e.g. their blackbody properties.  %\cite{DeLaRosa2017} compard the UV light curve of SNe~\2009ip and 2011ht to modeled lightcurves. 
To our knowledge, no systematic and planned survey of the early phase of SNe IIn in the UV has been performed so far. In this paper, we present a sample of 12 SNe IIn detected and observed by the Zwicky Transient Facility (ZTF) \citep{Bellm2019,Graham2019} and followed-up in the UV by the {\it Neil Gehrels Swift Observatory} (\textit{Swift}) space telescope \citep{Gehrels2004}, using the Swift's Ultraviolet/Optical Telescope (UVOT;\citealt{Roming2005, Poole2008, Breeveld2011}).

We present the aforementioned observations in \S 2. In \S 3, we present some analysis of these observations. \S 4 is dedicated to constraining the fraction of SNe IIn exploding into aspherical CSM. We summarize our main results in \S 5. 

\section{Observations and data reduction}
In this section, we present the ZTF and \textit{Swift} observations of the 12 SNe IIn of our sample. 

\subsection{Discovery}\label{sec:discovery}

All 12 SNe IIn were detected by the ZTF automatic pipeline as potential transients in the data from the ZTF camera mounted on the $1.2$\,m Samuel Oschin telescope (P48, \citealt{Rahmer2008}). A duty astronomer reviewing the ZTF alert stream \citep{Patterson2019} via the ZTF GROWTH Marshal \citep{Kasliwal2019}.The host galaxies $r$-band magnitudes, as well as the coordinates, redshift and distance modulus of all objects are summarized in Table~\ref{table:param}. The Milky Way extinction was deduced from \cite{Schlafly2011} using the extinction curves of \cite{Cardelli1989}. 

\subsection{Selection criterion}
Since the beginning of operation, ZTF has found several spectroscopically confirmed SNe IIn per month. However, applying the criterion for asphericity from \cite{Soumagnac2019} depends on our ability to measure the evolution of $r_{\rm BB}$ -- the effective blackbody radius -- at the time when the CSM is still optically thick and obstructing our view of any expanding material. We selected only SNe IIn which were spectroscopically confirmed while still on their rise. This selection criterion was motivated by two reasons (1) the spectrum of the SNe IIn in ths early phase is still well described by a blackbody spectrum (2) the rise of the optical light curve gives a better handle on the evolution of $r_{\rm BB}$ than the peak phase (3) we assumed that rising SNe IIn are young enough to allow us to take several $Swift$ observations and still be in the regime where expanding material has not reached optically thin areas of the CSM. The initial classification, within ZTF, was triggered as part of a variety of programs: the Redshift Completeness Factor (RCF; \citealt{Fremling2019}) program, the Census of the Local Universe program \citep{De2019}, the Superluminous Supernovae program, the Rapidely Evolving Transients program, the Science Validation program or the SNe IIn program. Some of these objects were first reported and classified by other surveys,  see Table 1 for details.

\begin{figure*}
\begin{center}
\includegraphics[scale=.55]{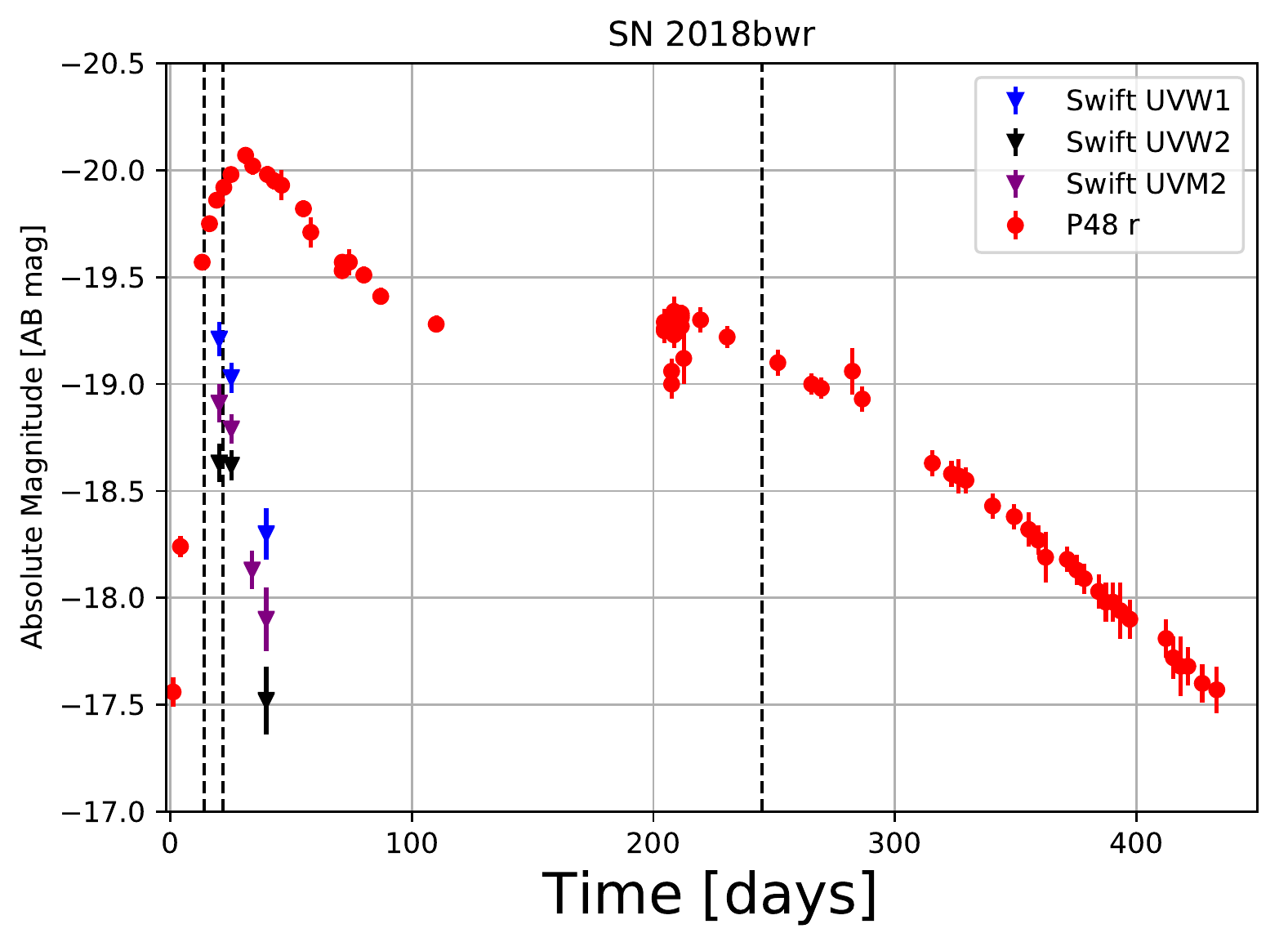}
\includegraphics[scale=.55]{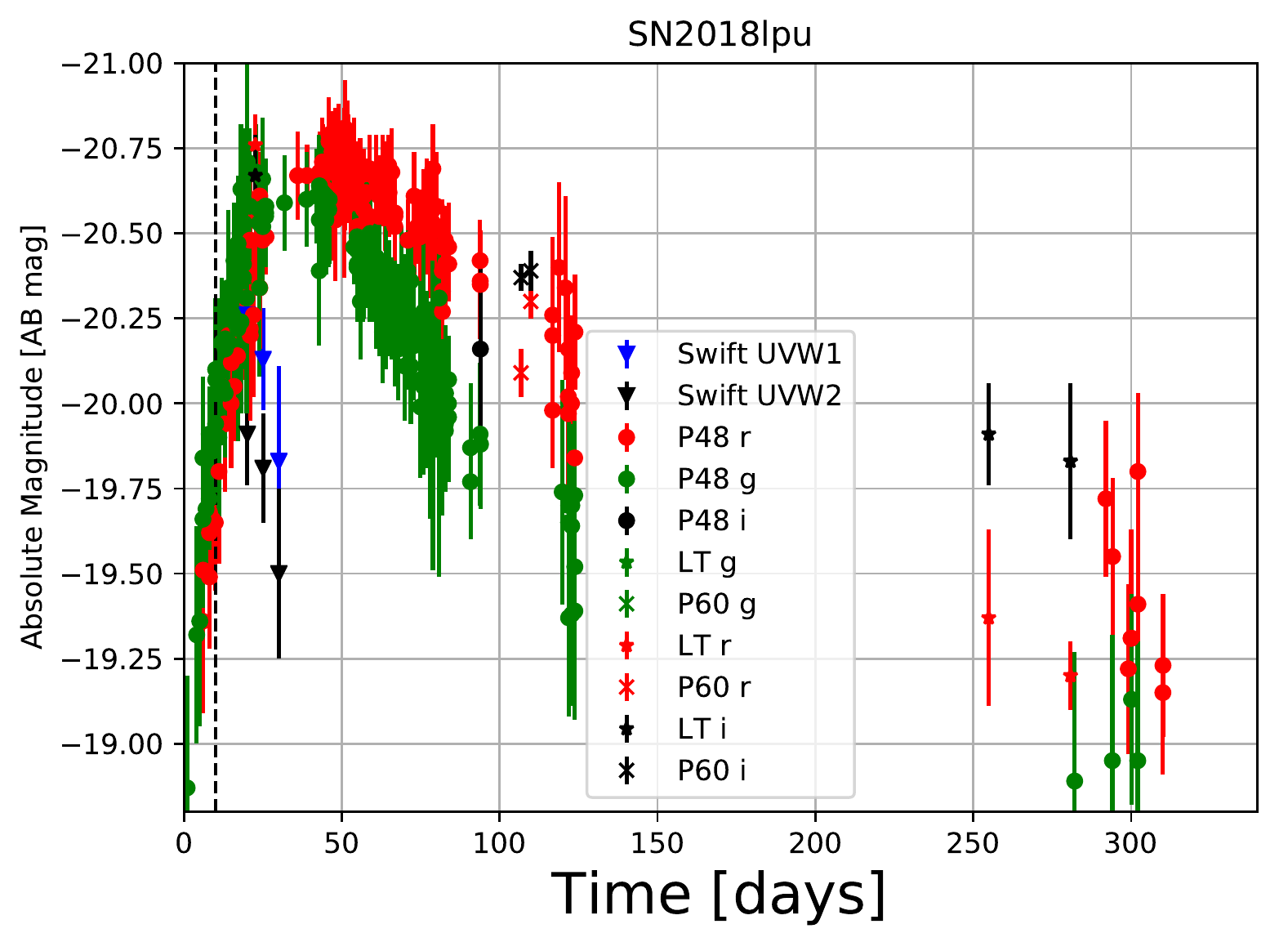}
\includegraphics[scale=.56]{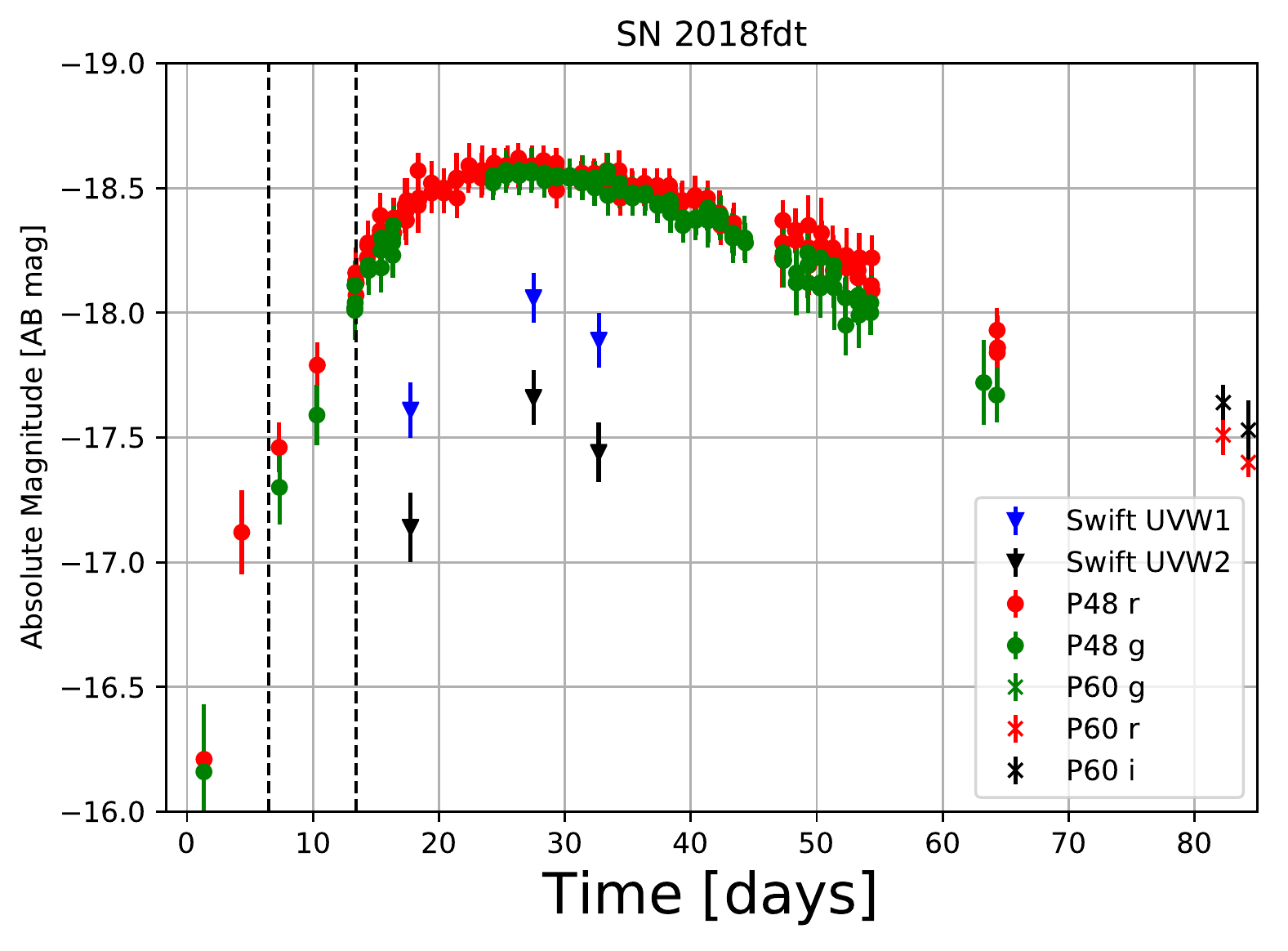}
\includegraphics[scale=.56]{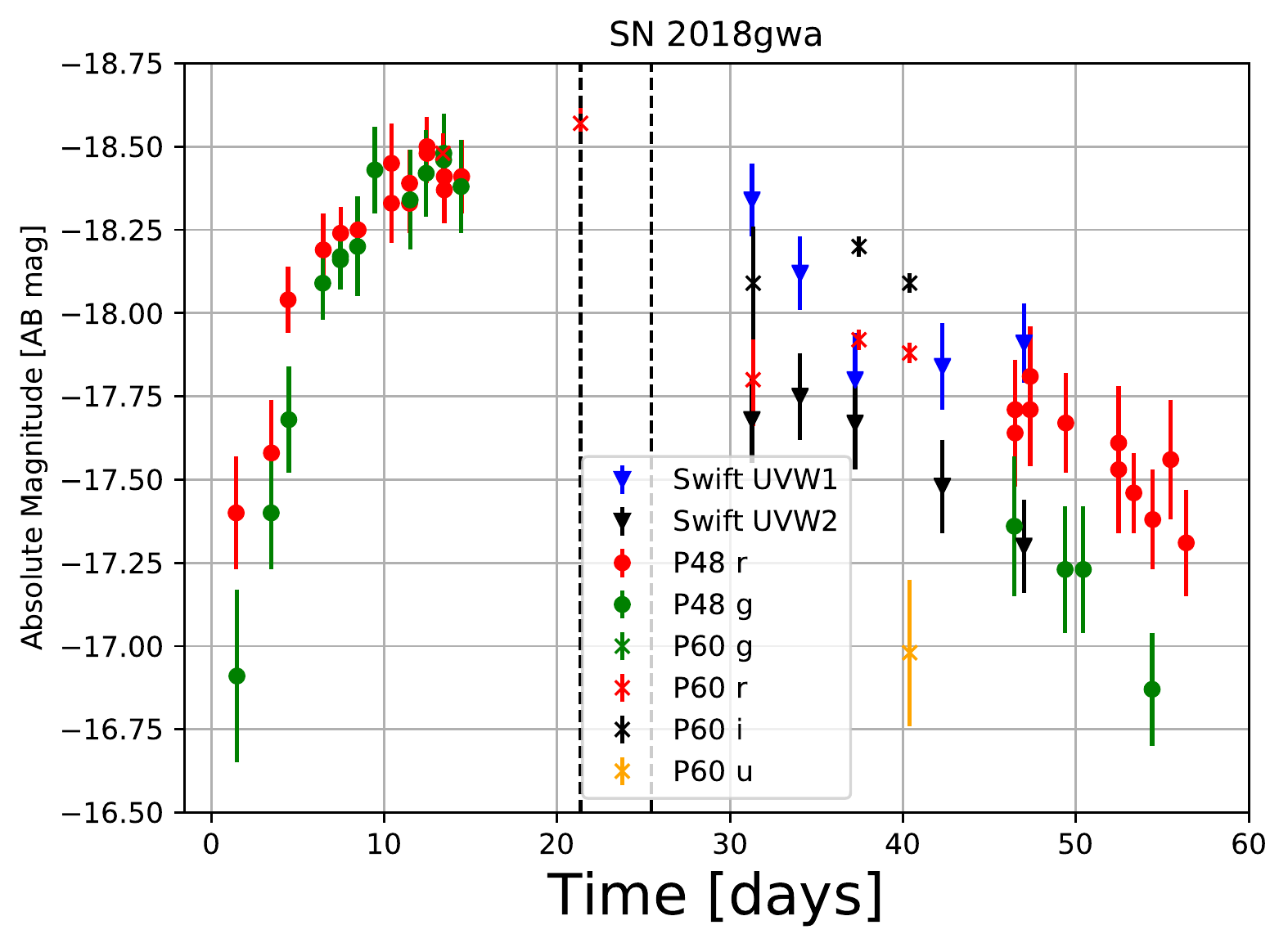}
\includegraphics[scale=.56]{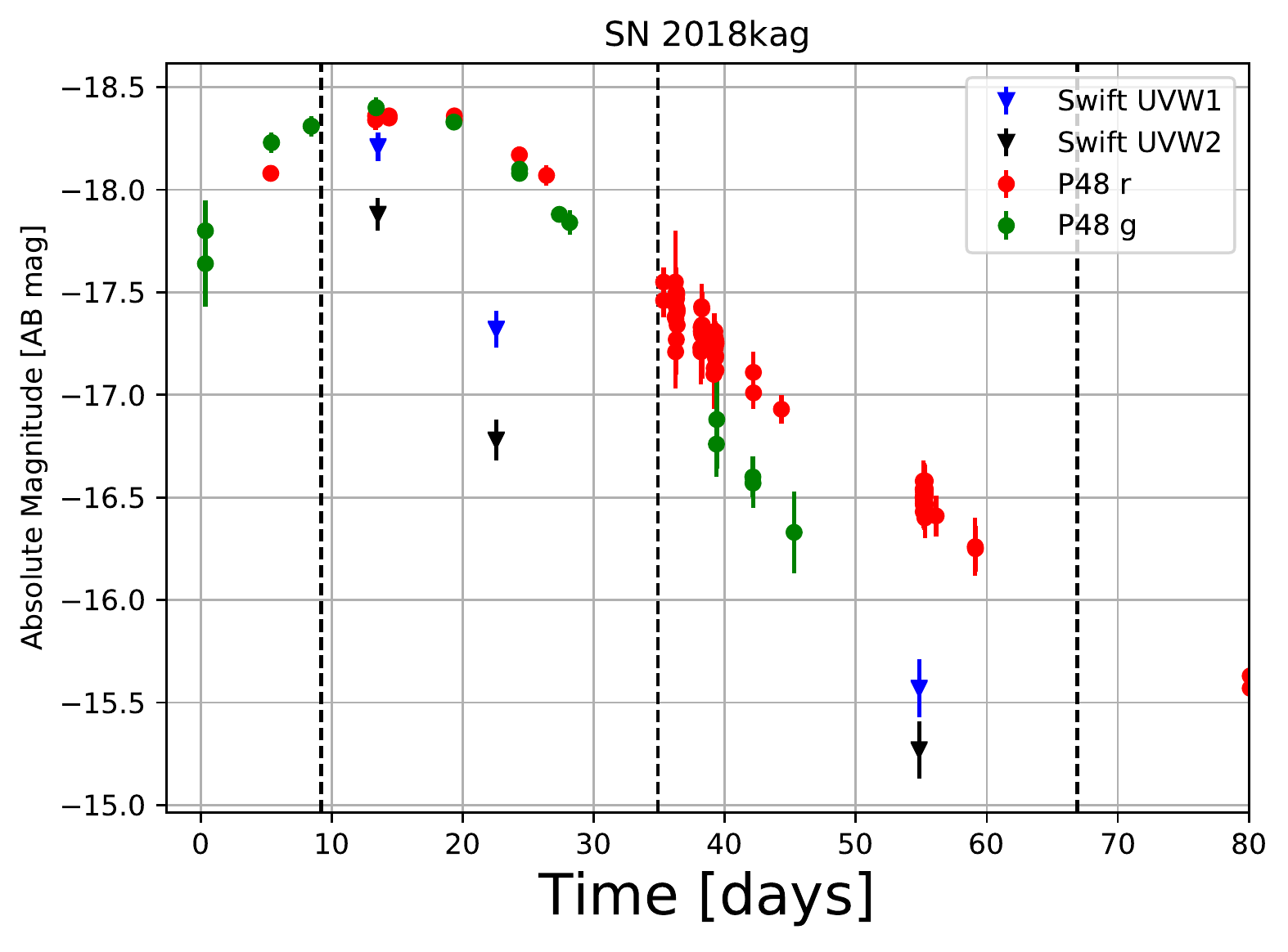}
\includegraphics[scale=.56]{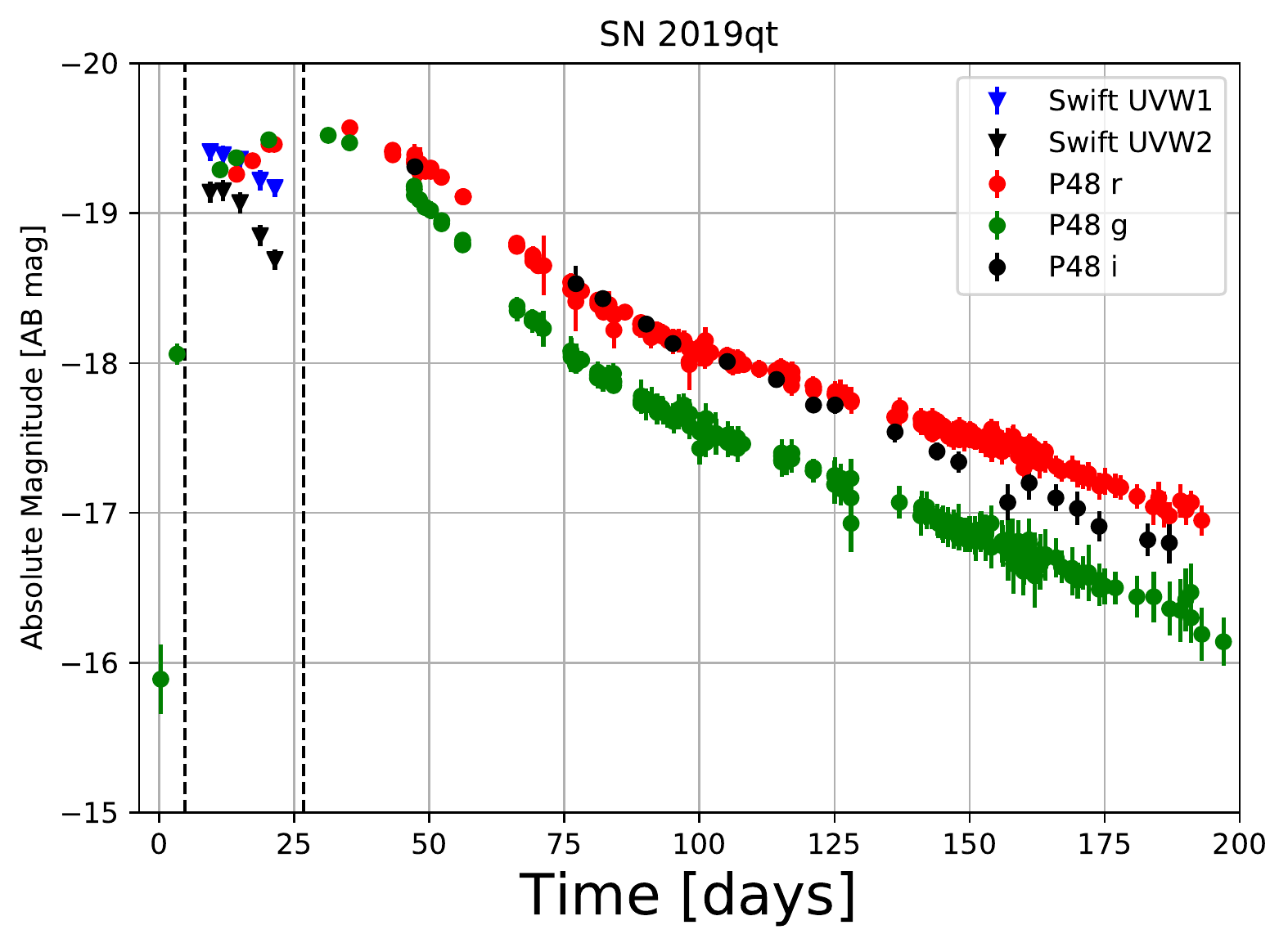}
%\includegraphics[scale=.35]{./2018_sub.png}
%\includegraphics[scale=.40]{./Screenshot_ds9.png}
%\caption{The light curve of all the objects in our sample. Time is shown relative to the estimated epoch at which the extrapolated light curve (Equation~\ref{eq:bolometric luminosity exp} and Equation~\ref{eq:bolometric luminosity exp}) is crossing zero: $t_{0}$, as derived in \S~\ref{sec:t0} and summarized in Table~\ref{table:param}. Black dashed lines indicate dates at which spectroscopic data exist.}
%\label{fig:lc1}
\end{center}
\end{figure*}

\begin{figure*}%\ContinuedFloat
\begin{center}
\includegraphics[scale=.56]{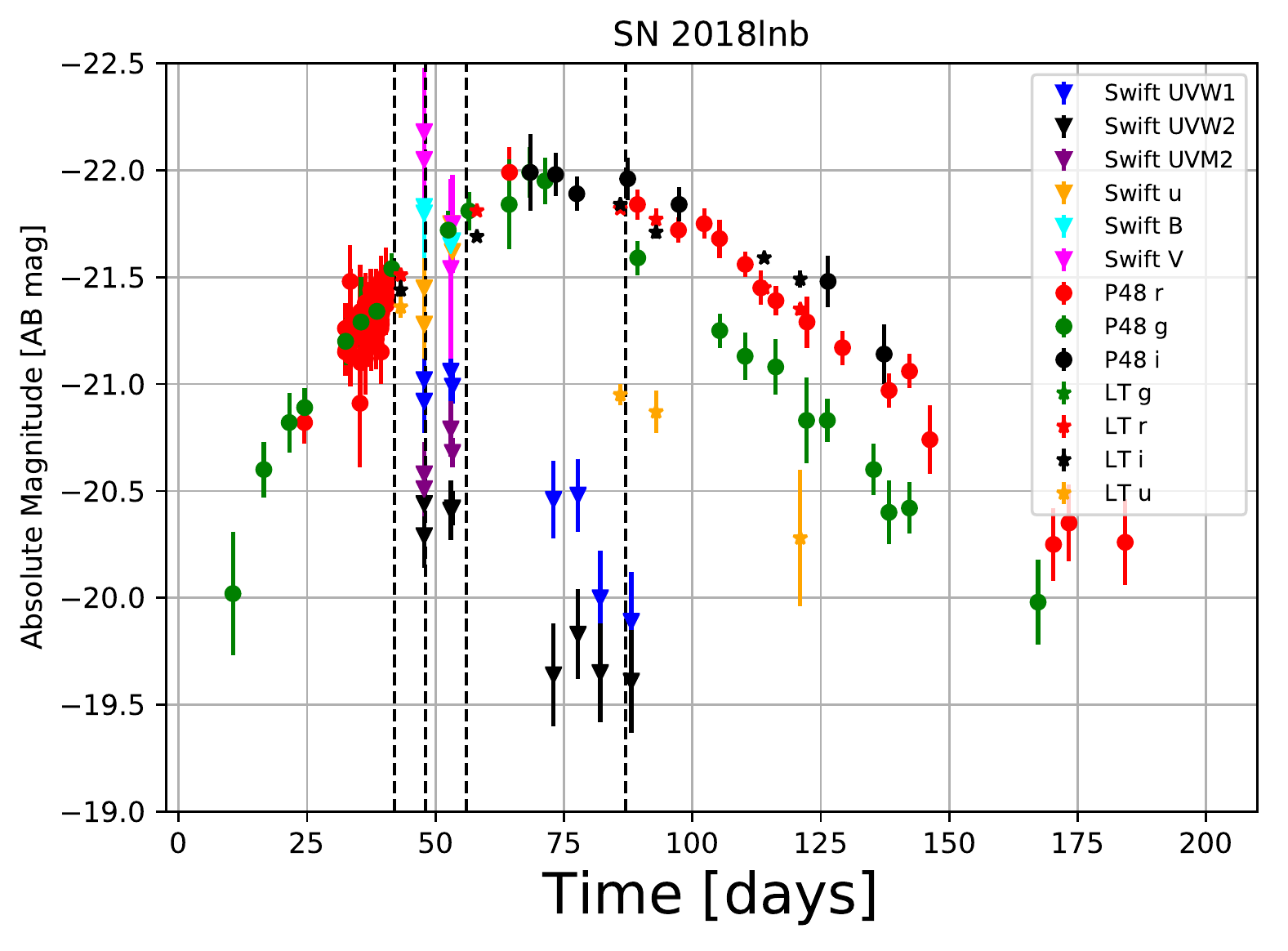}
\includegraphics[scale=.56]{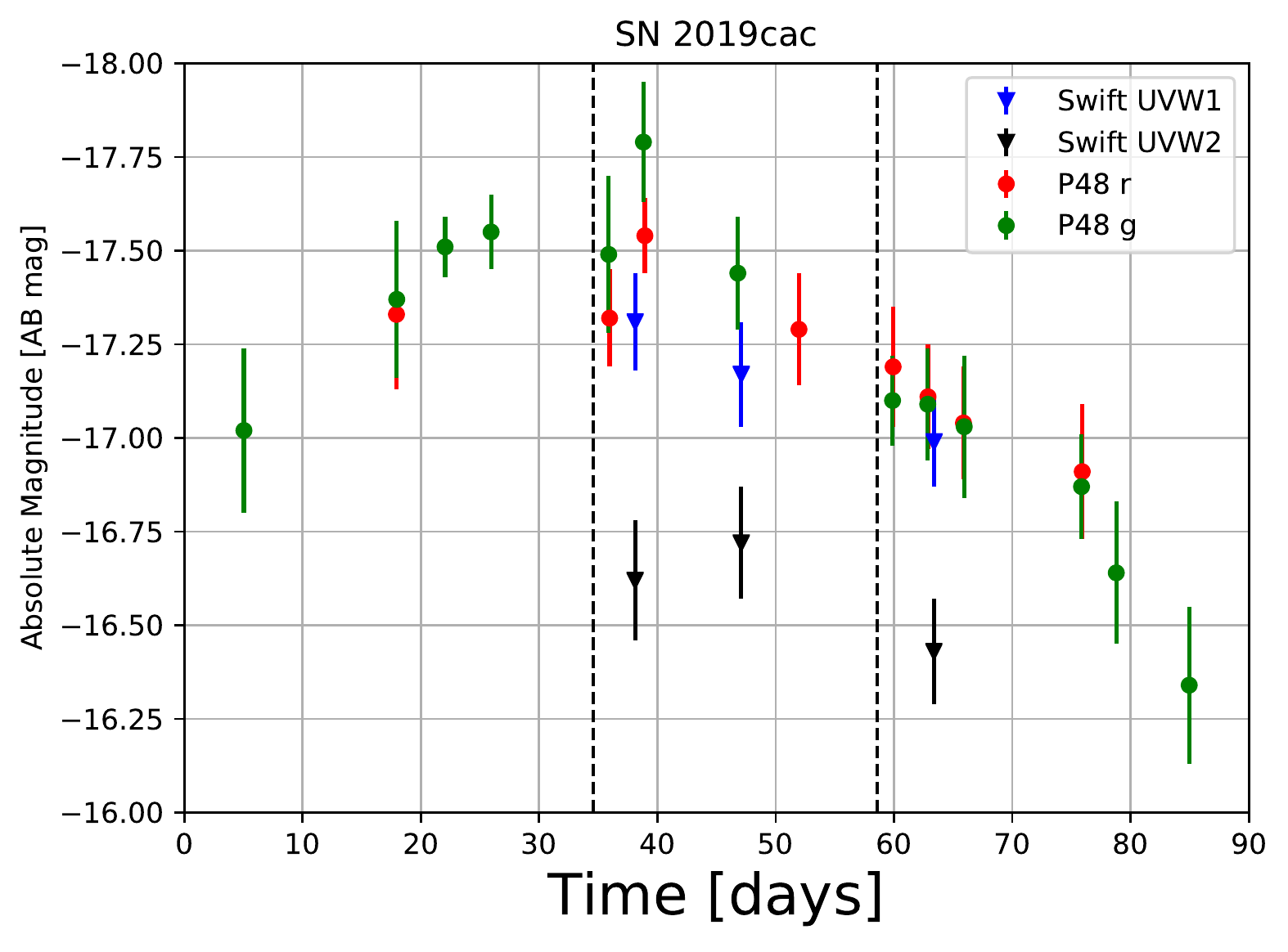}
\includegraphics[scale=.56]{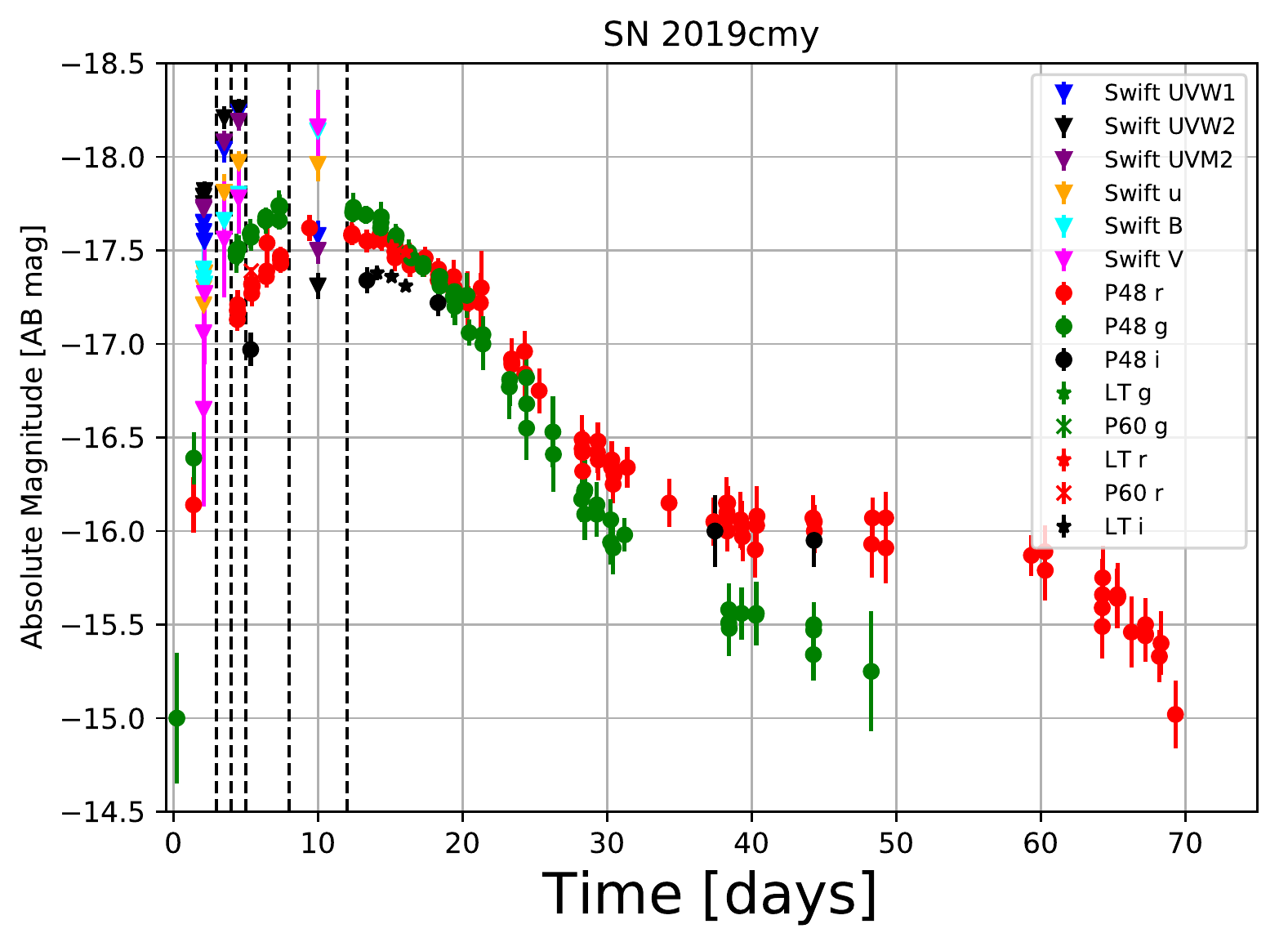}
\includegraphics[scale=.55]{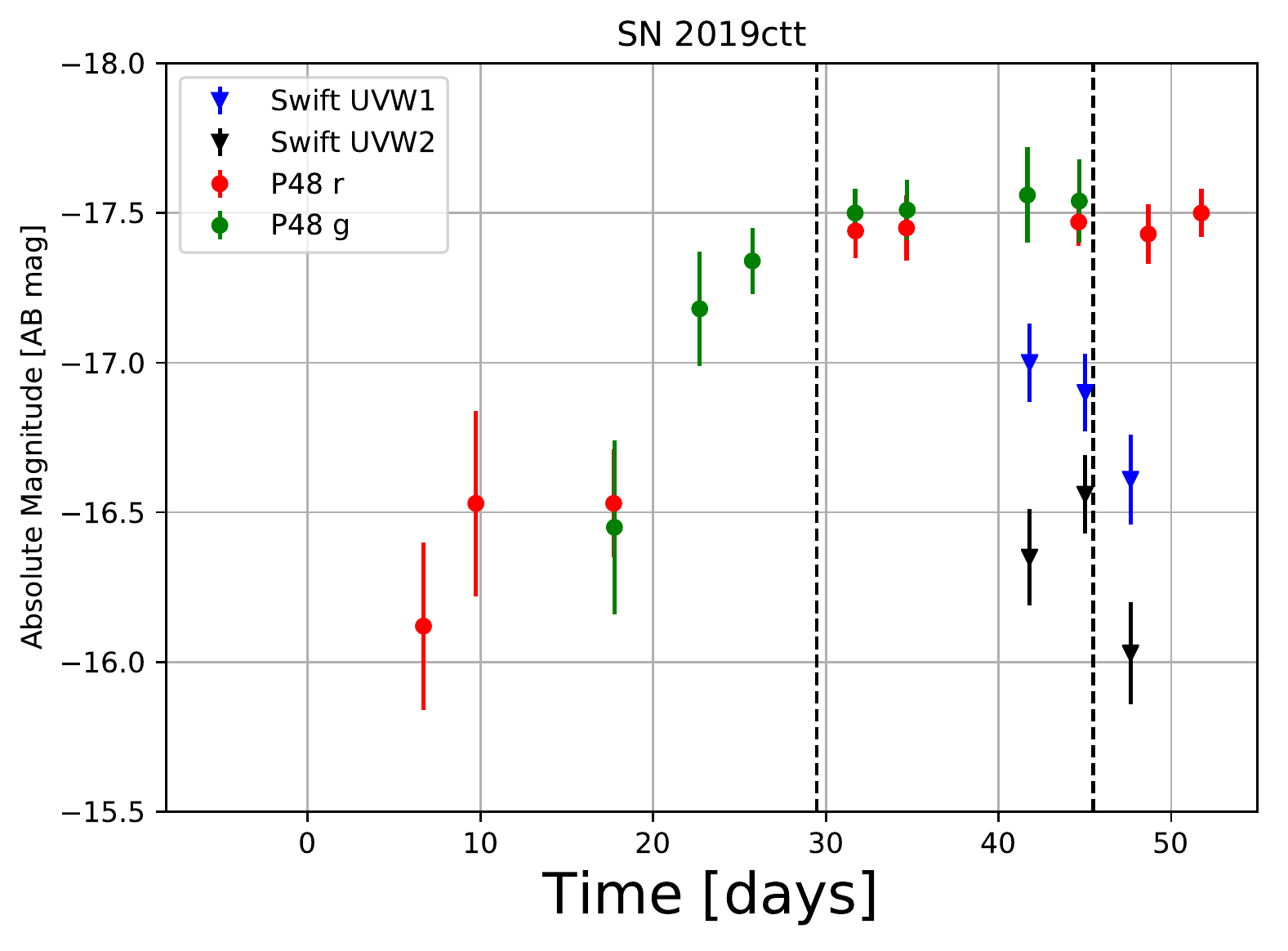}
\includegraphics[scale=.56]{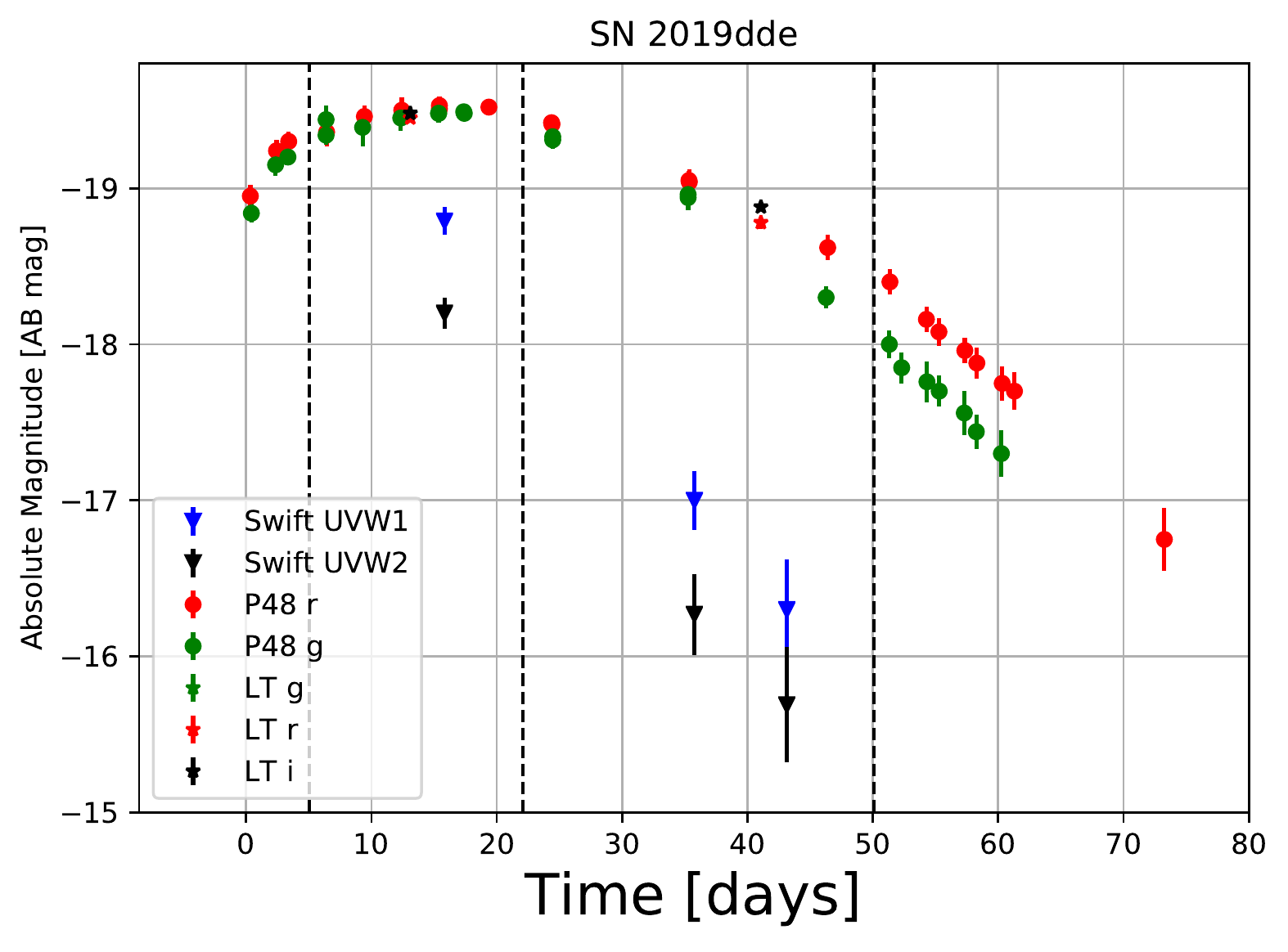}
\includegraphics[scale=.56]{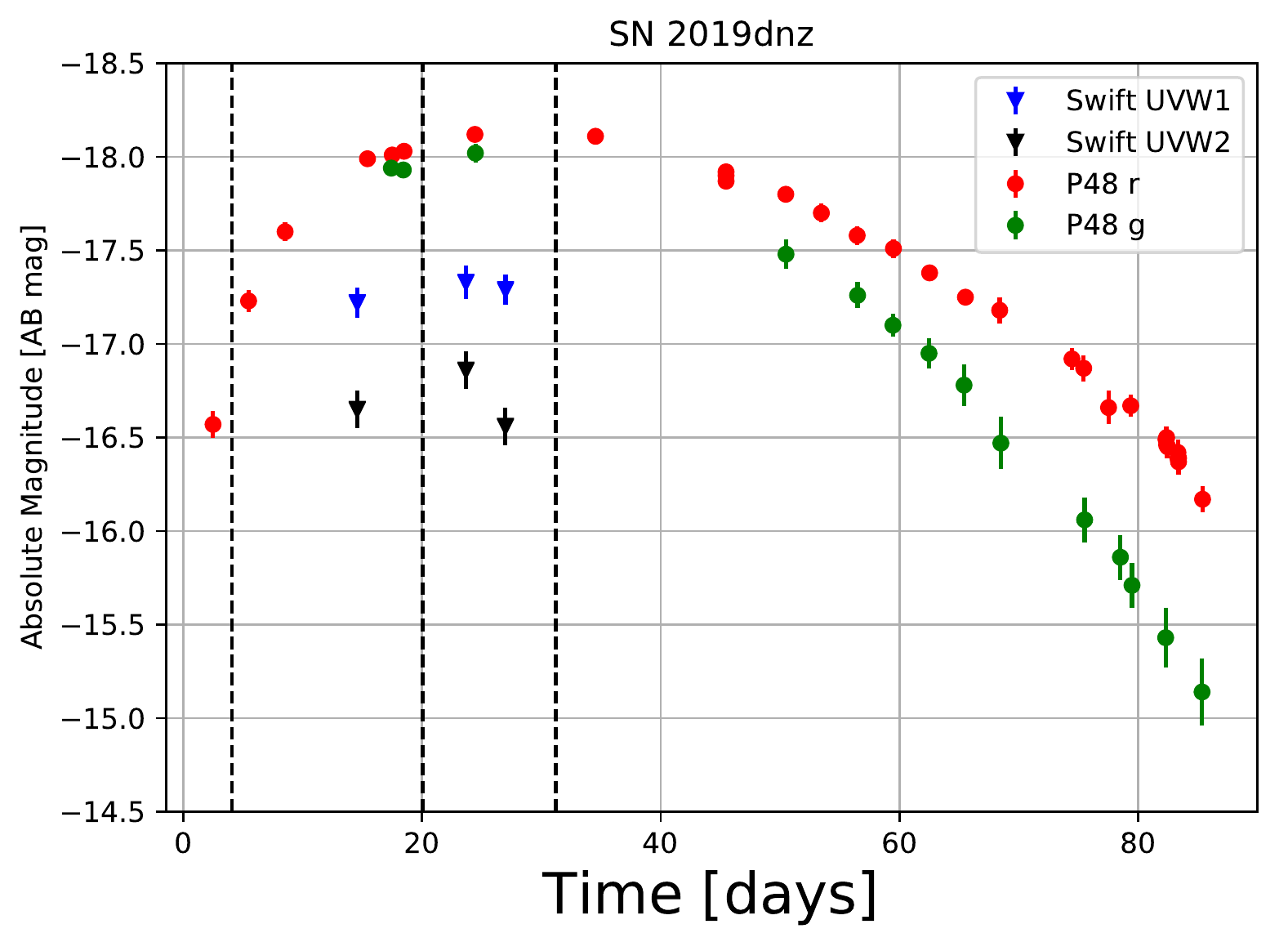}
\caption{The light curves of all the objects in our sample. Time is shown relative to the estimated epoch at which the extrapolated light curve (Equation~\ref{eq:bolometric luminosity exp} and Equation~\ref{eq:bolometric luminosity power}) is reaching zero: $t_{0}$, as derived in \S~\ref{sec:t0} and summarized in Table~\ref{table:t0}. The x-axis starts at the most recent non-detection, used as the lower limit of the prior in the $t_0$ fit. Black dashed lines indicate dates at which spectroscopic data exist.}
\label{fig:lc1}
\end{center}
\end{figure*}

\begin{deluxetable*}{lllllll}
\tablecolumns{7}
\tablecaption{Summary of observational parameters}
\tablewidth{0pt}
\tablehead{\colhead{IAU Name}&\colhead{ZTF Name}&\colhead{RA}&\colhead{Dec}&\colhead{Redshift}&\colhead{Distance modulus}&\colhead{$E_{B-V}$}\\%&\colhead{Reference}\\
%\colhead{} & \colhead{} & \colhead{} & \colhead{} & \colhead{} & %\colhead{modulus} &\colhead{extinction}\\%&\colhead{time}\\
\colhead{} & \colhead{} & \colhead{(deg)} & \colhead{(deg)} & \colhead{} & \colhead{(mag)} &\colhead{(mag)}}%&\colhead{($JD$)}}\\
\startdata
SN\,2018lpu & ZTF18abgrlpv&$283.937395$&$+47.441250$&$0.2104$&$40.10$&$0.055$\\%&$2458311.479$\\
SN\,2018fdt & ZTF18abltfho&$256.184755$&$+38.235567$&$0.055$&$36.91$&$0.036$\\%&$2458336.342$\\%
SN\,2018gwa & ZTF18abxbhov&$110.069724$&$+41.346650$&$0.0659$&$37.33$&$0.075$\\%&$2458377.109$\\ \hline 
SN\,2018bwr & ZTF18aavskep&$232.109019$&$+8.806157$&$0.046$&$36.50$&$0.036$\\%&$2458257.527$\\
SN\,2018kag & ZTF18acwzyor&$133.951981$&$+3.584153$&$0.02736$&$35.33$&$0.045$\\%&$2458467.593$\\
SN\,2019qt & ZTF19aadgimr&$224.794385$&$+43.819899$&$0.035$&$35.88$&$0.017$\\%&$2458491.587$\\
SN\,2018lnb & ZTF19aaadwfi & $159.583646$ & $+48.275291$ & $0.222$ & $40.23$&$0.012$\\%& $+48.2752905$ \\
SN\,2019cac & ZTF19aaksxgp & $207.5882959$ & $-2.506948$ & $0.0467$ & $36.53$ & $0.049$\\%&$-2.5069478$
SN\,2019cmy & ZTF19aanpcep & $227.211849$ & $+40.713750$ & $0.0314$ & $35.58$ & $0.015$\\%&$227.2118487$ \\+40.7137497
SN\,2019ctt & ZTF19aanfqug & $150.176198$ & $+12.039836$ & $0.0464$ & $36.50$ & $0.037$\\%& $2458554.219$\\
SN\,2019dde & ZTF19aaozsuh & $217.050160$ & $-1.580420$ & $0.06$ & $37.11$ & $0.052$\\%& $2458582.536$\\-1.5804196
SN\,2019dnz & ZTF19aaqasrq & $297.131153$ & $+2.913750$ & $0.025$ & $35.13$ & $0.183$\\%& $2458583.449$
\enddata
\tablecomments{The three first SNe are those for which we were unable to secure enough spectroscopic data in order to include them in our analysis of the CSM geometry (see \S~\ref{sec:analysis_spectroscopy}).
 SN\,2018lpu was discovered and classified by the ZTF survey; SN\,2018fdt was discovered by the ATLAS survey on 2018-08-14 as ATLAS18tuy \citep{SN2018fdt}, also detected by Gaia surveys as Gaia18chl, classified by ZTF \citep{SN2018fdt_class}; SN\,2018gwa was discovered \citep{SN2018gwa} and classified \citep{SN2018gwa_class} by ZTF, also detected by Gaia on 2018-10-05 as Gaia18cxl; The rest of the SNe in the table are all inclded in our analysis of the CSM geometry. SN\,2018bwr was discovered by the ATLAS survey on 2018-05-21 as ATLAS18ppb \citep{SN2018bwr}, also detected by PS1 and Gaia surveys as PS18aau and Gaia18bpl, classified by ZTF \citep{SN2018bwr_class}; SN\,2018kag was discovered by the ASAS-SN survey on 2018-12-17 as ASASSN-18abt and classified by \cite{SN2018kag_class}; SN\,2019qt was discovered \citep{SN2019qt} and classified \citep{SN2019qt_class} by ZTF, also detected by ATLAS, Gaia and PS1 as ATLAS19btl, Gaia19aid and PS19ahv; SN\,2018lnb was discovered and classified by ZTF \citep{SN2018lnb_class}; SN\,2019cac was discovered and classified by ZTF \citep{AT2019cac}, also detected by ATLAS and PS1 as ATLAS19doj and PS19ym; SN\,2019cmy was discovered \citep{SN2019cmy} and classified \citep{SN2019cmy_class} by ZTF, also detected by ATLAS as ATLAS19elx; SN\,2019ctt was discovered by ZTF \citep{SN2019ctt} and classified by SCAT \citep{SN2019ctt_class}; SN\,2019dnz was discovered by ZTF \citep{SN2019dnz} and classified by TCD \citep{SN2019dnz_class}, also detected by ATLAS as ATLAS19hra; SN\,2019dde was discovered by ZTF, classified by ZTF \citep{SN2019dde_class} and \citep{SOAR}, also detected by MASTER and PS1 as MASTER OT J142812.05-013615.2 and PS19aaa.}% The last column displays the time at which the extrapolated light curves (based on Equation~\ref{eq:bolometric luminosity exp} and Equation~\ref{eq:bolometric luminosity power}) are reaching zero, as derived in \S~\ref{sec:t0}. These times are used as an estimate of the explosion epoch.} %{\color{red}[add error \label{tab:format}}
\label{table:param}
\end{deluxetable*}

\subsection{Photometry}

%2018\,fif was observed in multiple bands for $\sim 2$ months. %, using \textit{GALEX} and two ground-based telescopes: the P48 telescope and the P60 telescope \citep{Law2009,Rau2009}.
%The SN was monitored during a rising phase ($t<36$ days) and a decay phase ($t>243$ days) but not around peak luminosity. 
All the light curves are shown in Figure~\ref{fig:lc1}. The photometry is reported in electronic Table~\ref{table:photo} and is available via WISeREP\footnote{https://wiserep.weizmann.ac.il}.

%\textit{Swift} observations of the 2018\,fif field started on 2018 August 21 and 11 observations were obtained with a cadence of $\sim1$ day. 
%The \textit{Swift} data reduction was done using the $UVOTsource$\footnote{?} tool. 
Photometry was obtained using the ZTF camera mounted on the P48 telescope, through the P48 $r$ and $g$ filters. Data were obtained with a cadence of about $1-3$ days, to a limiting AB magnitude of $\rm{r}=20.5$ mag and $\rm{g}=21$ mag. The P48 data were automatically reduced using the ZTF pipeline \citep{Masci2019}, using the image subtraction algorithm {\tt ZOGY} by \cite{Zackay2016}.
%For the data reduction of the P48 data, we used the FPipe pipeline \citep{Fremling2016}. 

The robotic $1.52$\,m telescope at Palomar (P60; \citealt{Cenko2016}) was used with a $2048\times2048$-pixel "Rainbow" CCD camera \citep{Benami2012,Blagorodnova2018} and $g'$, $r'$, $i'$ SDSS filters. Data reduction of the P60 data was performed using the FPipe pipeline \citep{Fremling2016}, using the image subtraction algorithm by \cite{Zackay2016}.

For several SNe IIn discussed in this work, We acquired multi-band images with the optical imager (IO:O) on the Liverpool Telescope \citep[LT;][]{ssr+04} . Images reductions were provided by the basic IO:O pipeline and image subtraction was performed versus PS1 ($g’$,$r’$,$I’$,$’z$-bands) or SDSS ($u’$-band) reference imaging, following the techniques of \citet{Fremling2016}.  PSF photometry was performed relative to PS1/SDSS photometric standards.

The $Swift$ UVOT data were retrieved from the NASA Swift Data Archive\footnote{\href{https://heasarc.gsfc.nasa.gov/cgi-bin/W3Browse/swift.pl}{ https://heasarc.gsfc.nasa.gov/cgi-bin/W3Browse/swift.pl }} and reduced using standard software distributed with HEAsoft version 6.26\footnote{\href{https://heasarc.nasa.gov/lheasoft/}{ https://heasarc.nasa.gov/lheasoft/}}. Photometry was measured using the FTOOLSs {\tt uvotimsum} and {\tt uvotsource} with a 5’’ circular aperture. 

None of the SNe IIn in our sample were detected with the  $Swift$ XRT camera. %To remove the host contribution, we obtained and coadded two final epoch in all broad-band filters and built a host template using uvotimsum and uvotsource with the same aperture used for the transient. 

%We calibrated the P60 data in the following way. The $r$-band light curve was scaled so that its average value during the time window covered by both telescopes matches the average value of the P48 $R$-band photometric data. The $g$-band and $i$-band data were scaled to match the synthetic photometry of the calibrated spectroscopic data (\S~\ref{sec:obs-spectroscopy}). The synthetic photometry used for the calibration and for other purposes in this paper was computed with the PyPhot\footnote{http://mfouesneau.github.io/docs/pyphot/} pipeline (Fouesneau, in preparation).
%Although the photometric data available for 2018\,fif do not cover the peak, the data during the rise and decay allow to place an upper limit on the absolute magnitude at peak: with  $M_r\lesssim-20$, 2018\,fif is at the bright-end of the observed SNe IIn, together with e.g., SN 2006gy \citep{Ofek2007,Smith2007}, SNe 2008fq \citep{Thrasher2008,Taddia2013} or SN\,2003ma \citep{Rest2009,Rest2011}. In particular, it is brighter than all SNe in the sample by \cite{Kiewe2012}, which was designed to be unbiased.

\begin{deluxetable*}{llllllllll}
\tablecolumns{10}
\tablecaption{Photometry}
\tablewidth{0pt}
\tablehead{\colhead{Object}&\colhead{Epoch}&\colhead{Mag}&\colhead{Magerr}&\colhead{Flux}&\colhead{Abs. mag}&\colhead{Abs. magerr}&\colhead{Filter}&\colhead{Instrument}\\
\colhead{}&\colhead{(JD)}&\colhead{(AB)}&\colhead{(AB)}&\colhead{(erg/s)}&\colhead{(AB)}&\colhead{(AB)}&\colhead{}&\colhead{}}
\startdata
%\tablehead{\colhead{Object}&\colhead{Epoch}&\colhead{Mag}&\colhead{Magerr}&\colhead{Flux}&\colhead{Fluxerr}&\colhead{Abs. mag}&\colhead{Abs. magerr}&\colhead{Filter}&\colhead{Instrument}\\
%\colhead{}&\colhead{(JD)}&\colhead{(AB)}&\colhead{(AB)}&\colhead{(erg/s)}&\colhead{(erg/s)}&\colhead{(AB)}&\colhead{(AB)}&\colhead{}&\colhead{}}
%ZTF18aavskep&$2458273.8166$&$16.75$&$0.01$&$5.267e-16$&$0.049e-16$&$-19.75$&$0.01$&$r$&ZTF+P48\\
%ZTF19aadgimr&$2458502.9868$&$16.59$&$0.04$&$1.097e-15$&$0.040e-15$&$-19.29$&$0.04$&$g$&ZTF+P48\\
%ZTF19aadgimr&$2458586.8067$&$17.75$&$0.04$&$1.366e-16$&$0.050e-16$&$-18.13$&$0.04$&$i$&ZTF+P48\\
%ZTF18aavskep&$2458277.8361$&$17.87$&$0.09$&$1.833e-15$&$0.152e-15$&$-18.63$&$0.09$&$UVW2$&Swift+UVOT\\
%ZTF18aavskep&$2458277.8383$&$17.58$&$0.09$&$2.004e-15$&$0.166e-15$&$-18.91$&$0.09$&$UVM2$&Swift+UVOT\\
%ZTF18aavskep&$2458277.8405$&$17.29$&$0.08$&$1.984e-15$&$0.146e-15$&$-19.21$&$0.08$&$UVW1$&Swift+UVOT\\
ZTF18aavskep&$2458273.8166$&$16.75$&$0.01$&$(5.267\pm0.049)\times10^{-16}$&$-19.75$&$0.01$&$r$&ZTF+P48\\
ZTF19aadgimr&$2458502.9868$&$16.59$&$0.04$&$(1.097\pm0.040)\times10^{-15}$&$-19.29$&$0.04$&$g$&ZTF+P48\\
ZTF19aadgimr&$2458586.8067$&$17.75$&$0.04$&$(1.366\pm0.050)\times10^{-16}$&$-18.13$&$0.04$&$i$&ZTF+P48\\
ZTF18aavskep&$2458277.8361$&$17.87$&$0.09$&$(1.833\pm0.152)\times10^{-15}$&$-18.63$&$0.09$&$UVW2$&Swift+UVOT\\
ZTF18aavskep&$2458277.8383$&$17.58$&$0.09$&$(2.004\pm0.166)\times10^{-15}$&$-18.91$&$0.09$&$UVM2$&Swift+UVOT\\
ZTF18aavskep&$2458277.8405$&$17.29$&$0.08$&$(1.984\pm0.146)\times10^{-15}$&$-19.21$&$0.08$&$UVW1$&Swift+UVOT\\
\enddata
\tablecomments{This table is available in its entirety in machine-readable format in the online journal. A portion is shown here for guidance regarding its form and content. Time is shown relative to the estimated epoch at which the extrapolated light curve (based on Equation~\ref{eq:bolometric luminosity power} and Equation~\ref{eq:bolometric luminosity exp}) is reaching zero, as derived in \S~\ref{sec:t0} and shown in Table~\ref{table:param}.}% To compute the apparent magnitudes from the counts, the zero-point for the $nUV$ data is $ZP_{nUV}=20.08$ and the zero-point for the P48 data is $ZP_{P48}=27.00$.}
\label{table:photo}
\end{deluxetable*}

%\begin{figure*}
%\begin{center}
%\includegraphics[scale=.80]{/Users/maayanesoumagnac/PostDoc/projects/2018fif/pdf_files/LC_absmag.pdf}
%\caption{The light curve of 2018\,fif. Time is shown relative to the estimated epoch at which the extrapolated light curve (Equation~\ref{eq:bolometric luminosity}) is crossing zero: $t_{0}=$ (2018 August 21), as derived in \S~\ref{sec:bolo}. Black dashed lines indicate dates at which spectroscopic data exist.} 
%\label{fig:lc}
%\end{center}
%\end{figure*}

\subsection{Spectroscopy}\label{sec:obs-spectroscopy}

Optical spectra of all SNe were obtained using the telescopes and spectrographs listed in Table~\ref{table:obs}. The spectra were used to determine the redshift from the narrow host lines (H${\rm \alpha}$). All the spectra were corrected for Galactic extinction as deduced from \cite{Schlafly2011}, using \cite{Cardelli1989} extinction curves. 

\begin{figure*}
\begin{center}
\includegraphics[scale=.35]{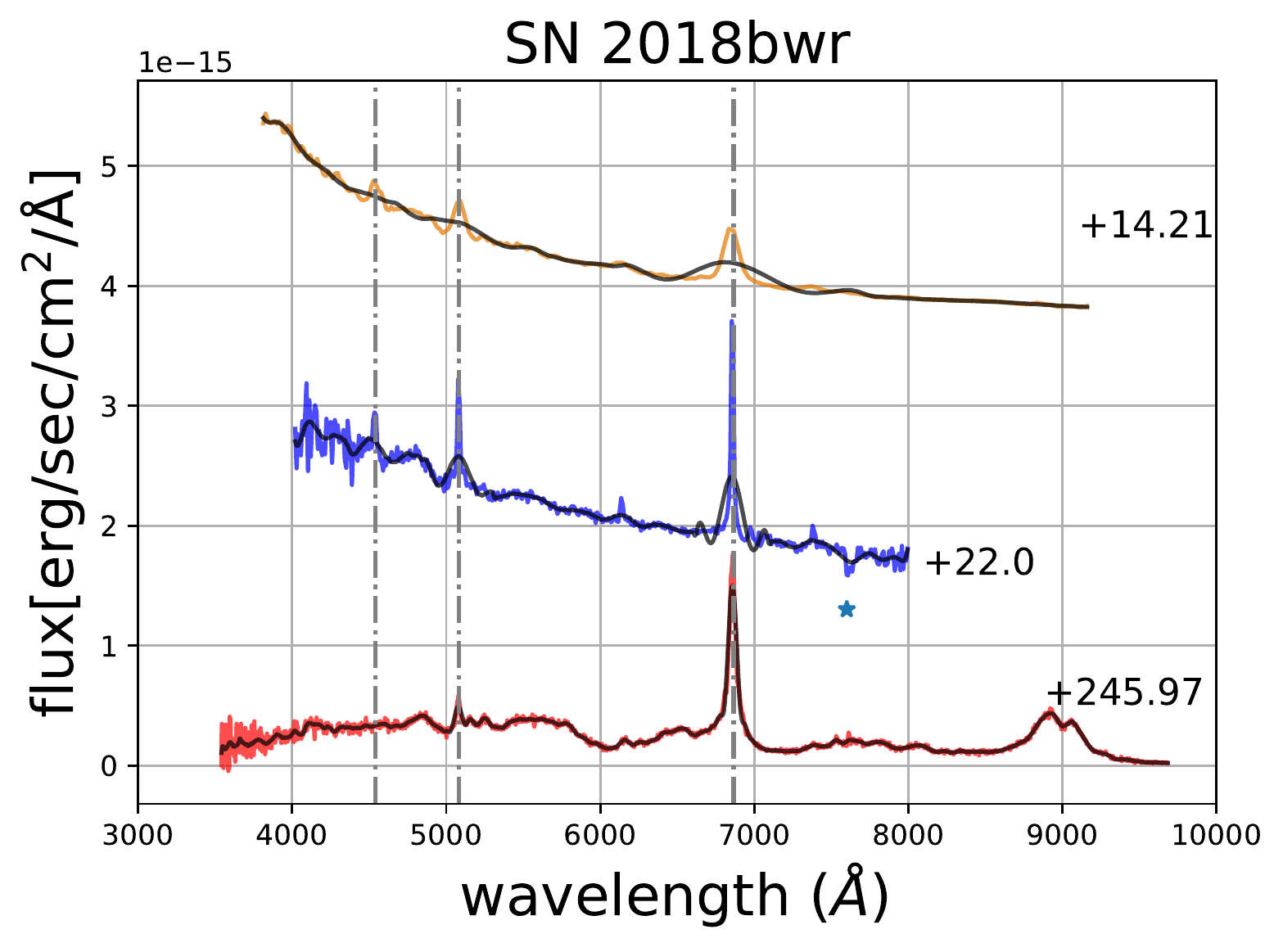}
\includegraphics[scale=.35]{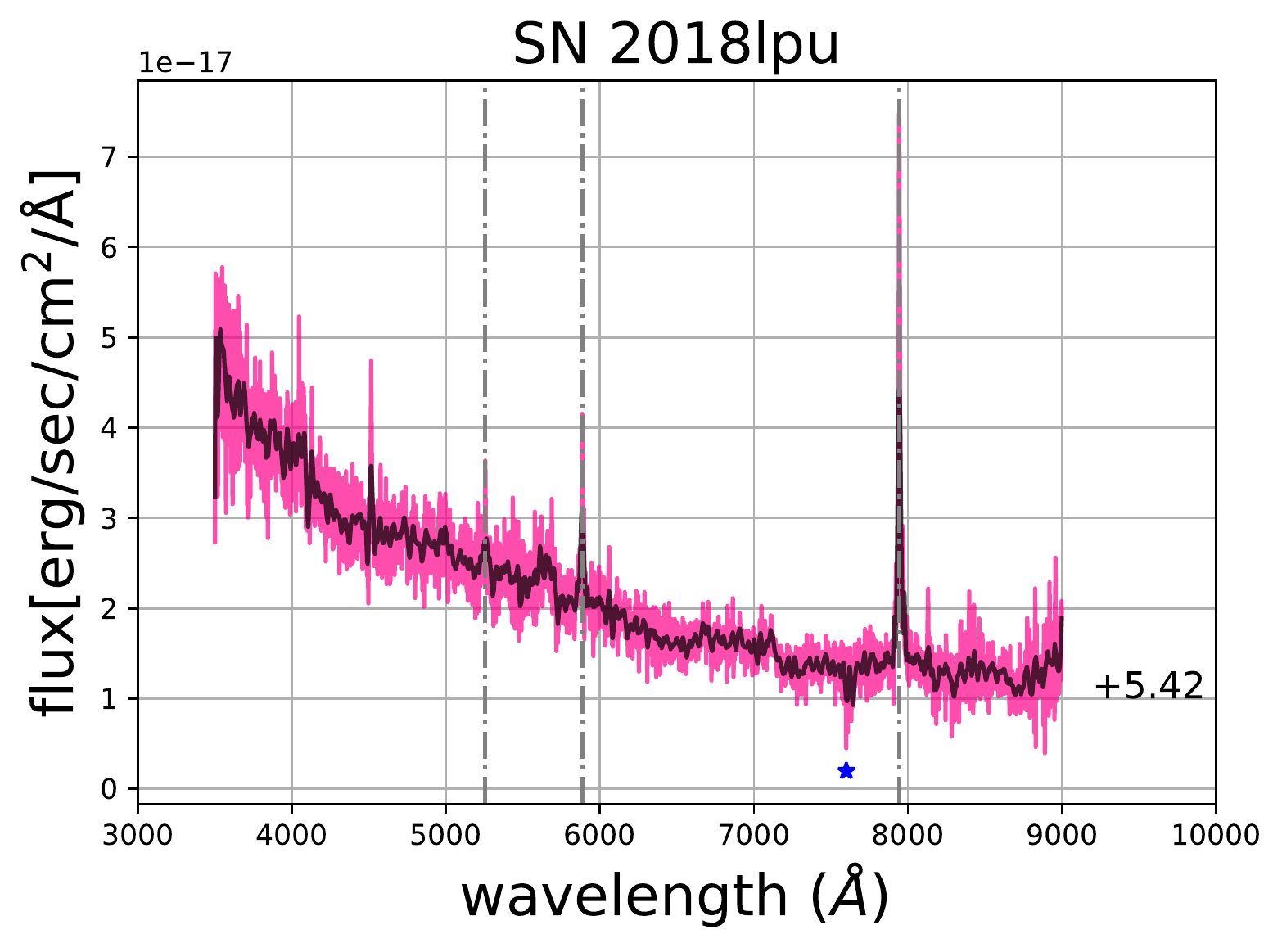}
\includegraphics[scale=.35]{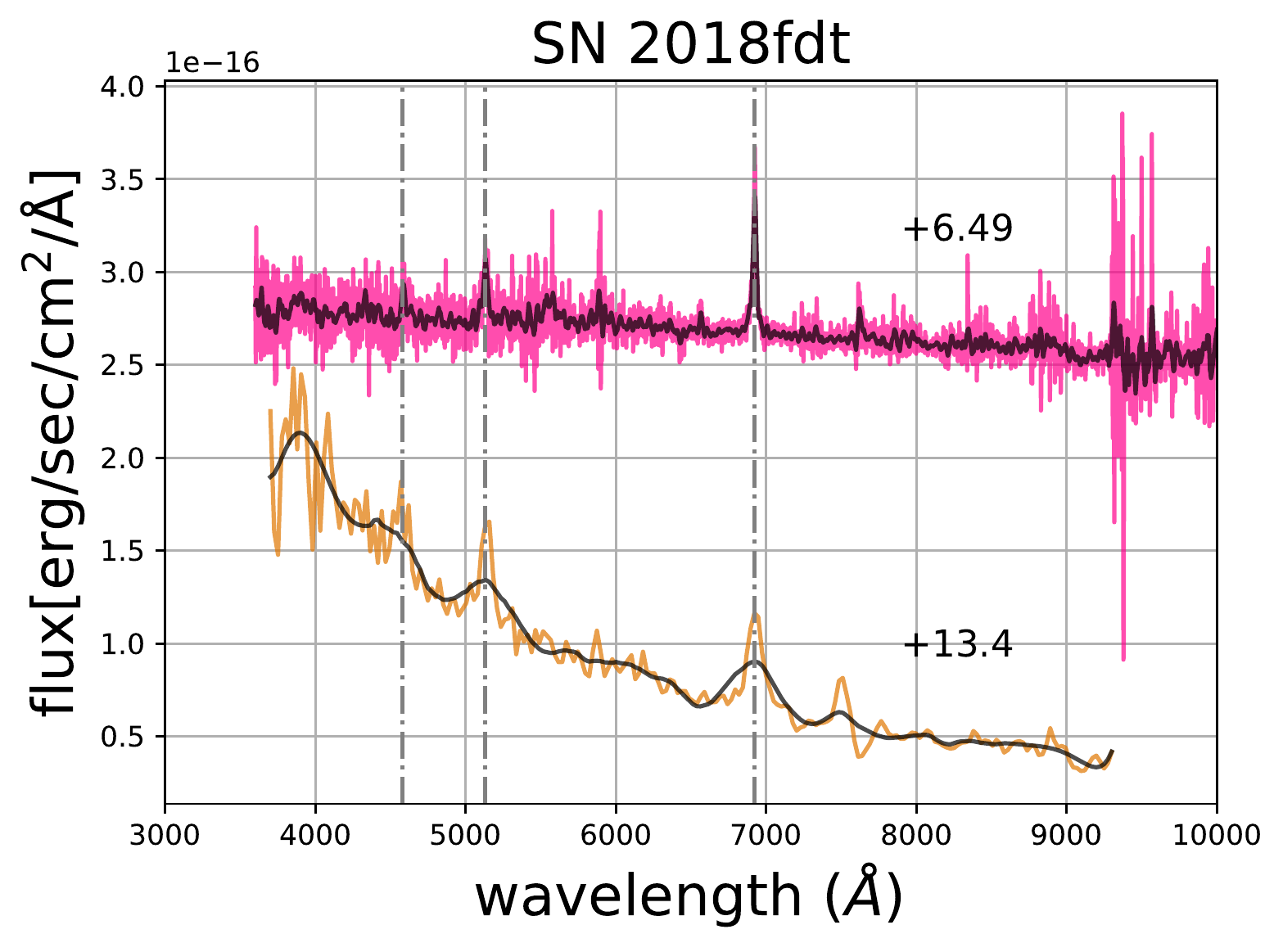}
\includegraphics[scale=.35]{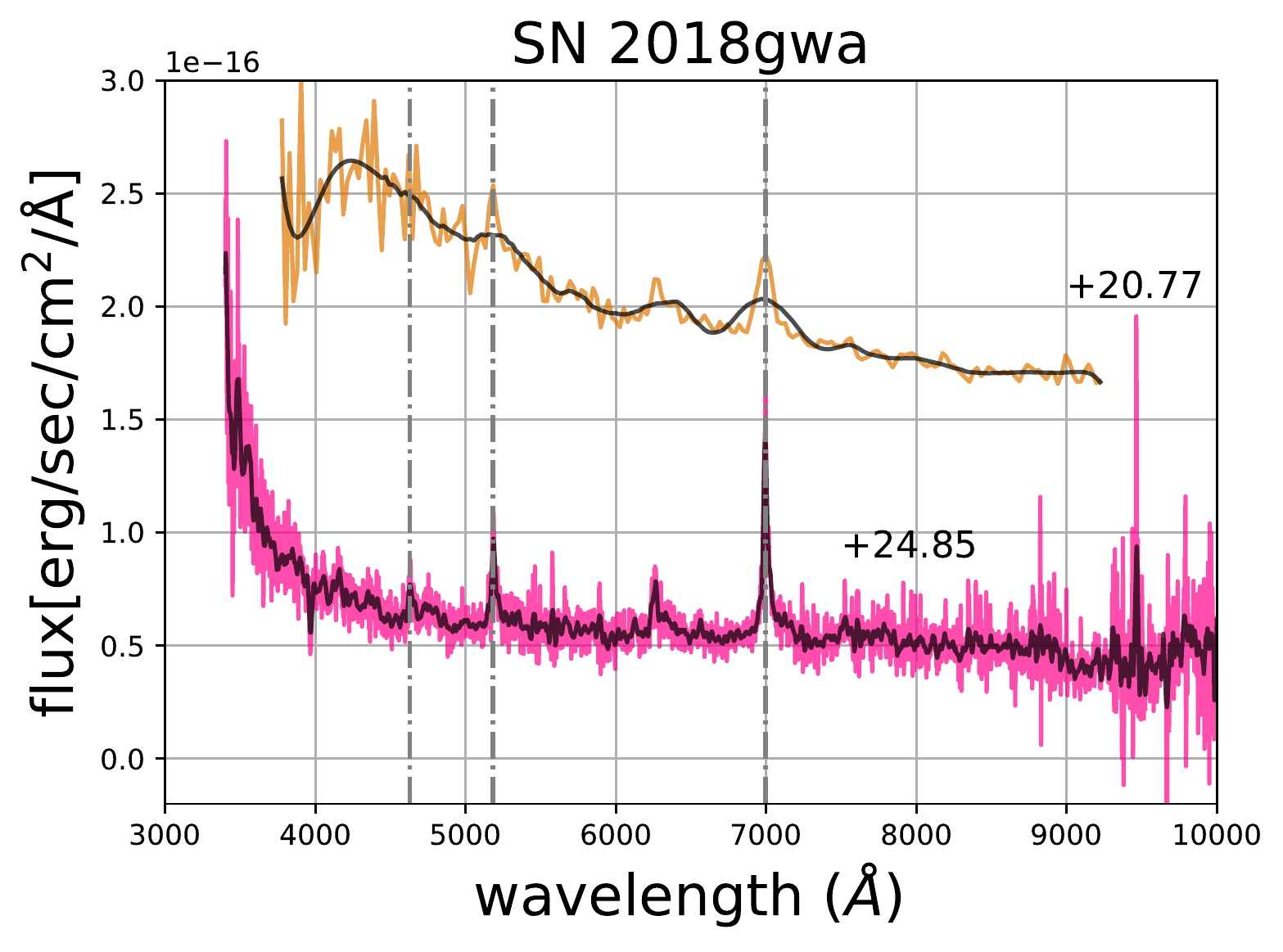}
\includegraphics[scale=.35]{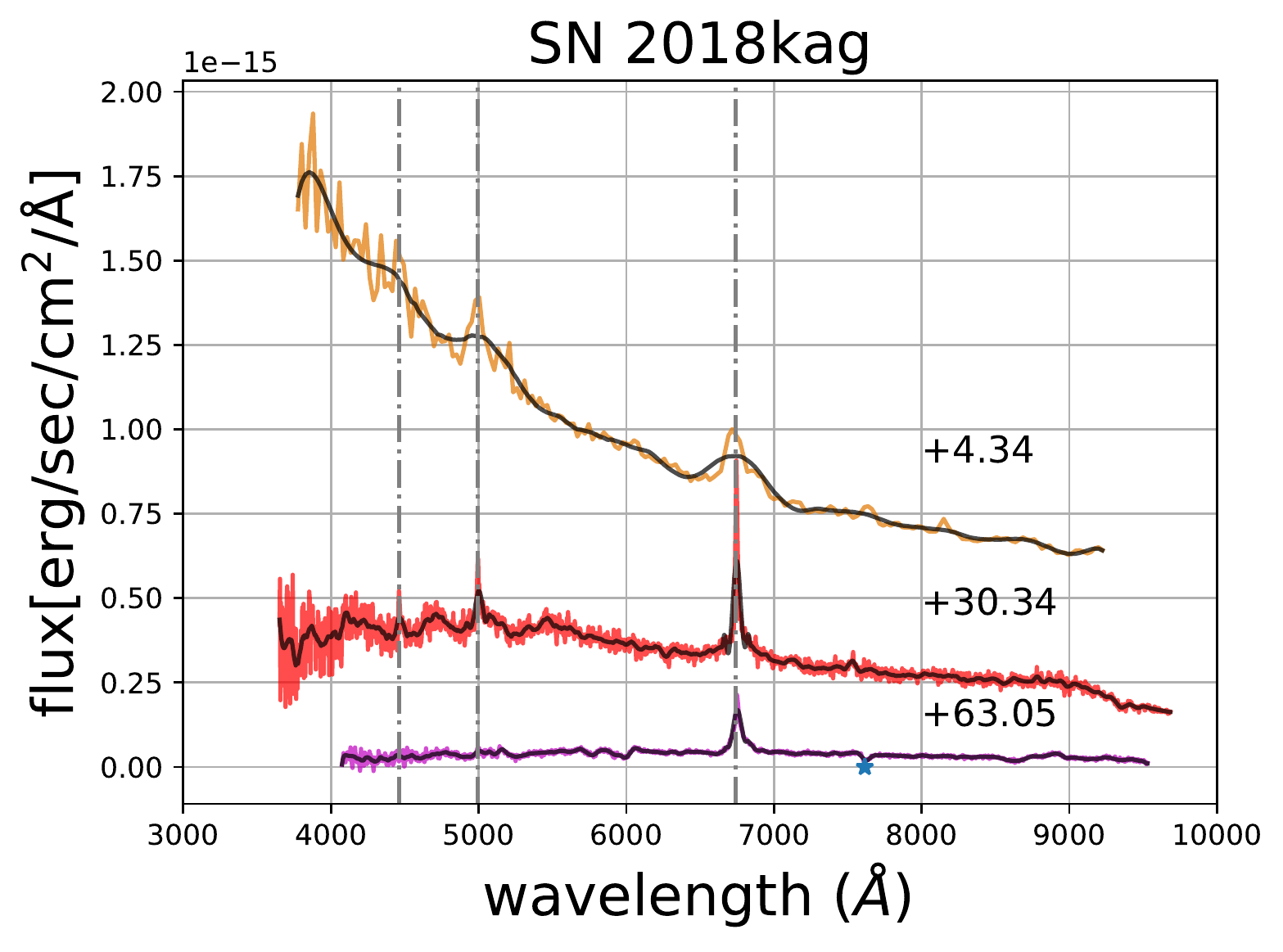}
\includegraphics[scale=.35]{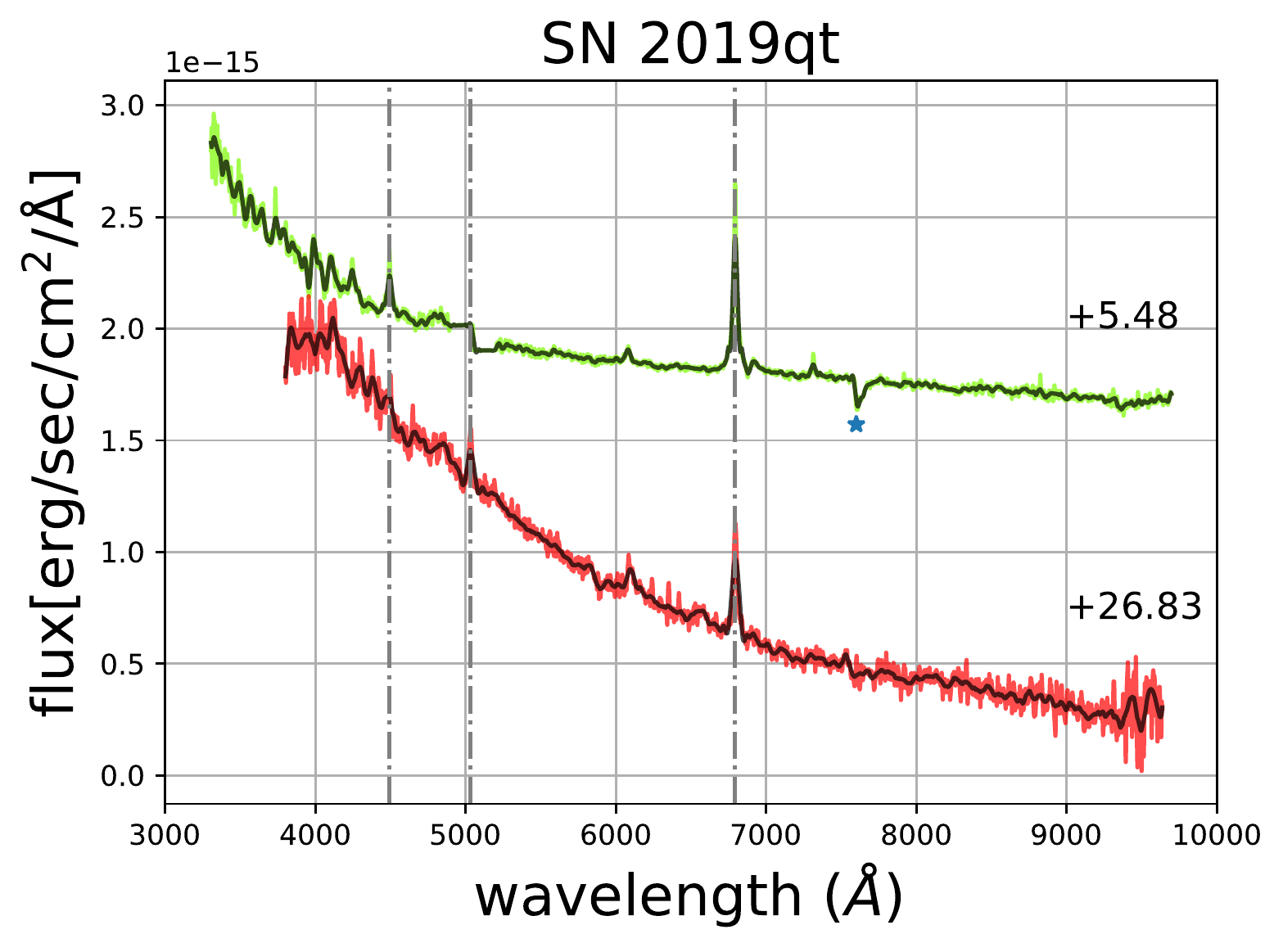}
\includegraphics[scale=.35]{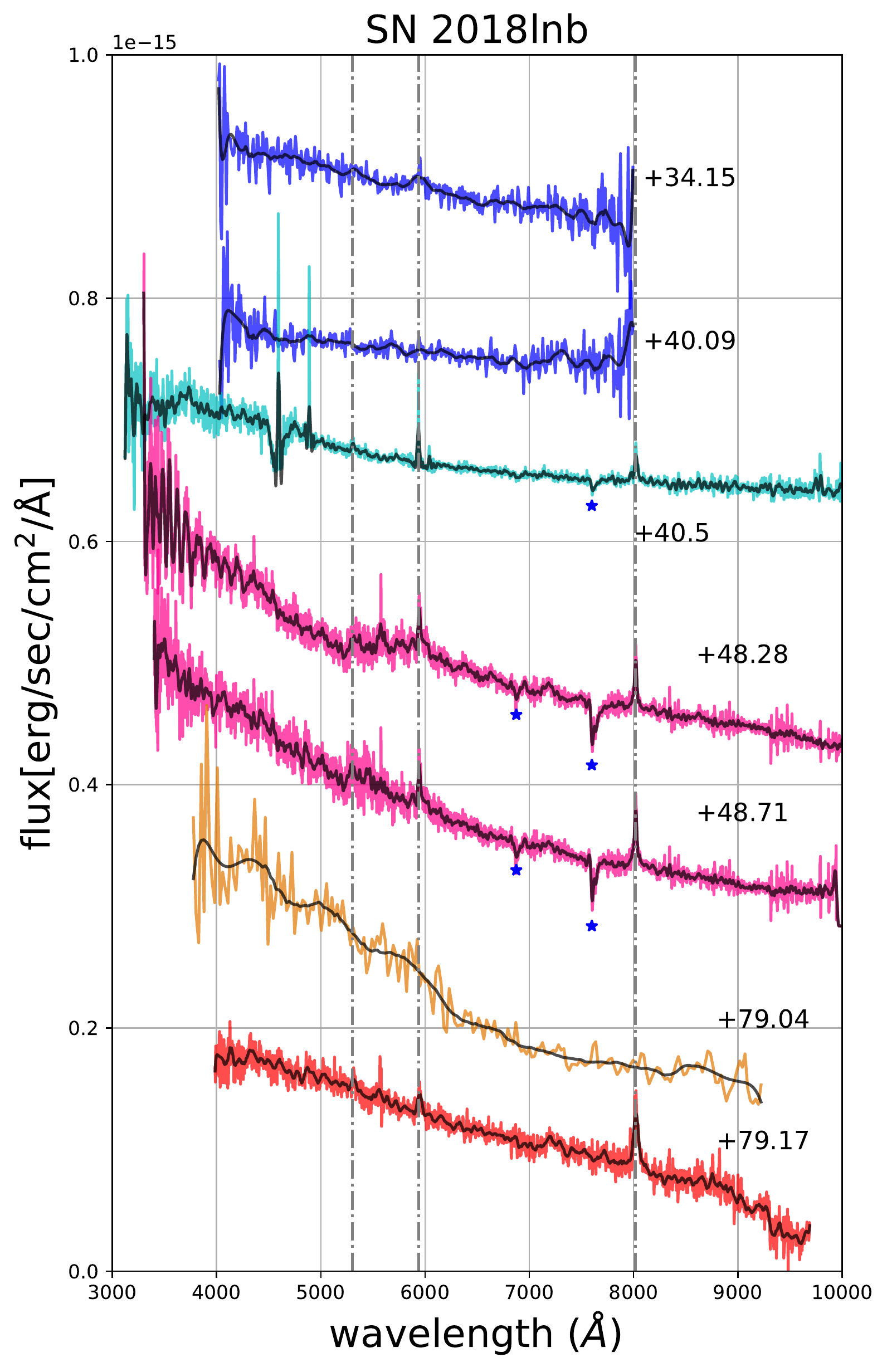}
\includegraphics[scale=.35]{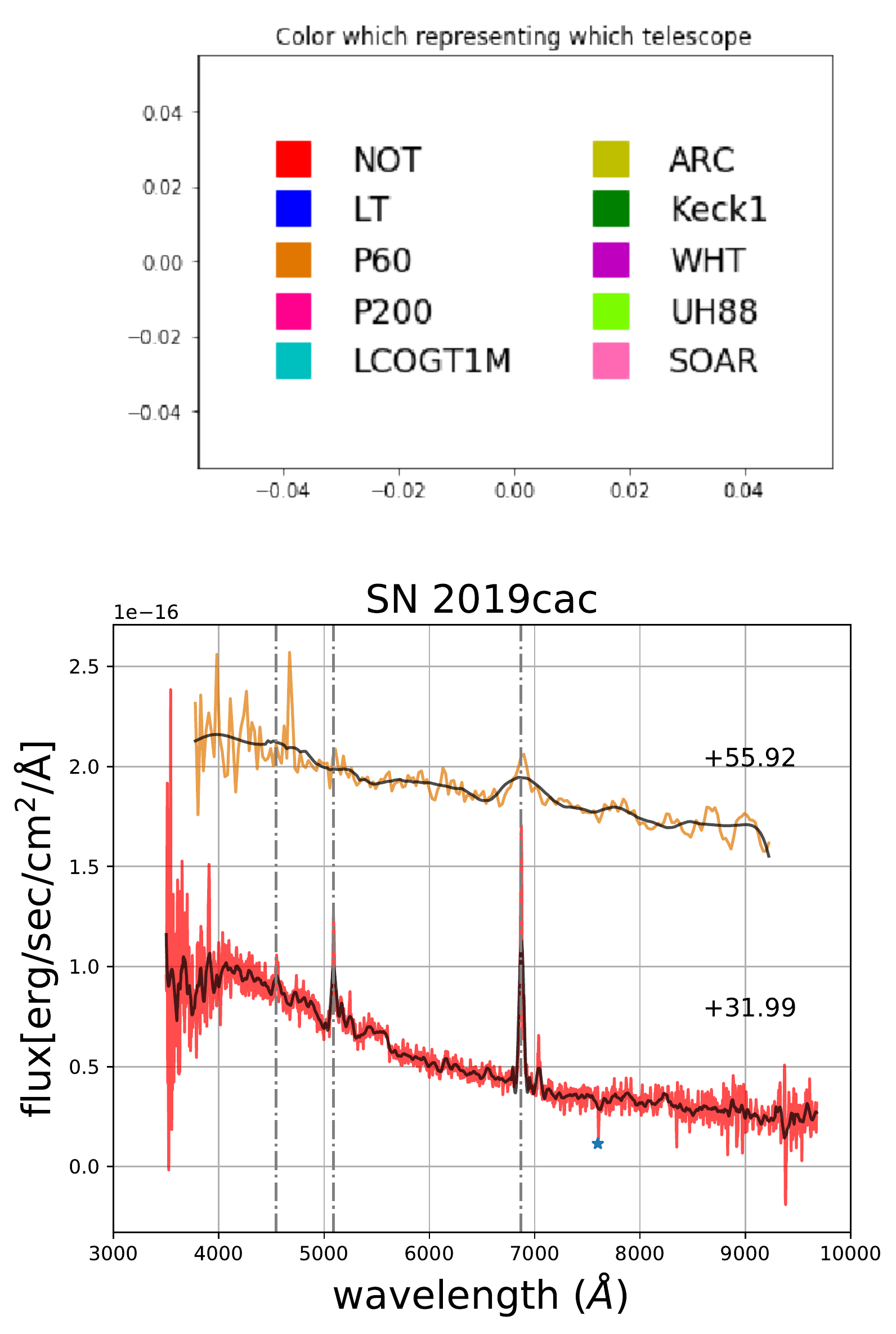}
\includegraphics[scale=.35]{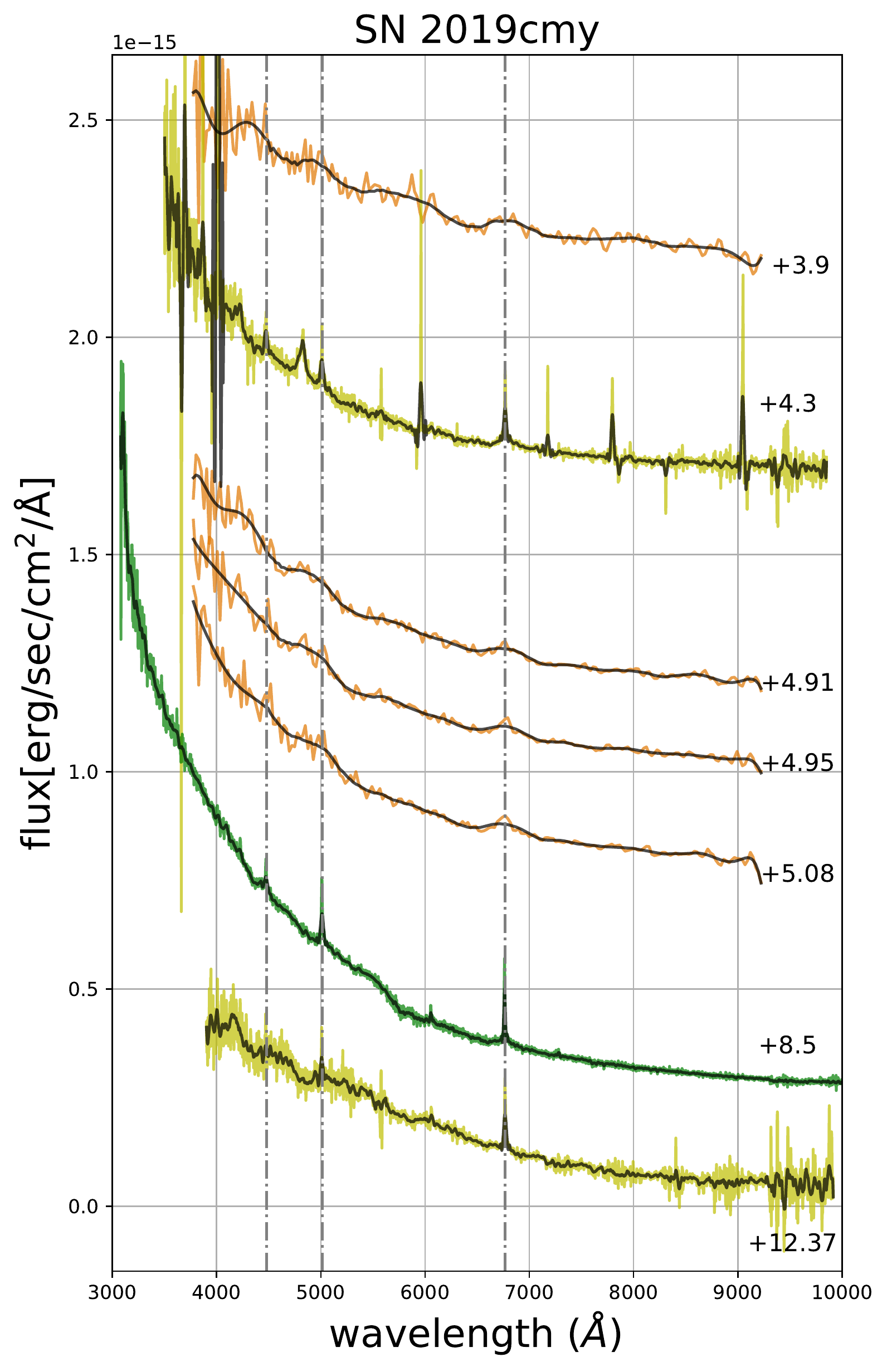}
\includegraphics[scale=.35]{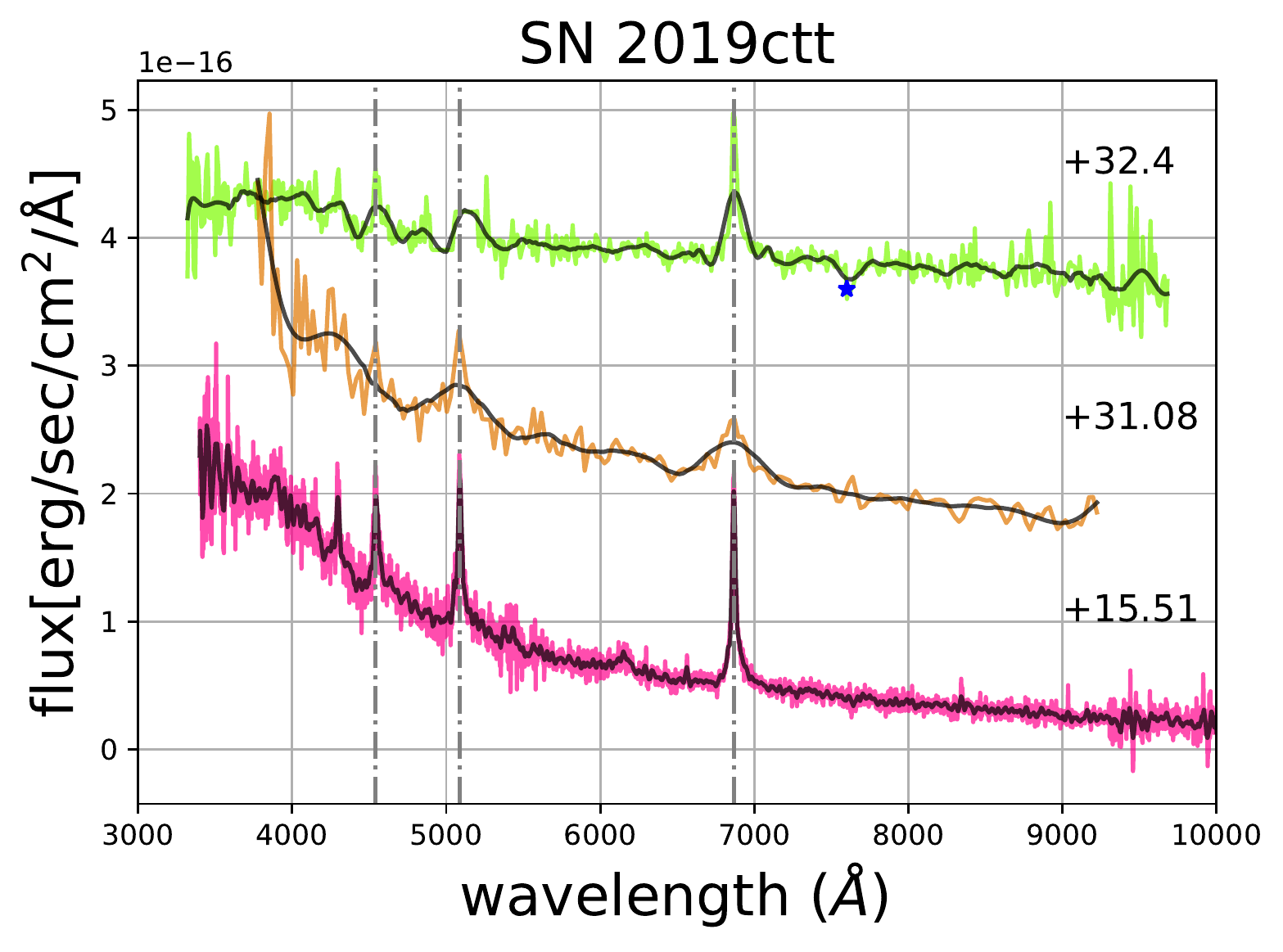}
\includegraphics[scale=.35]{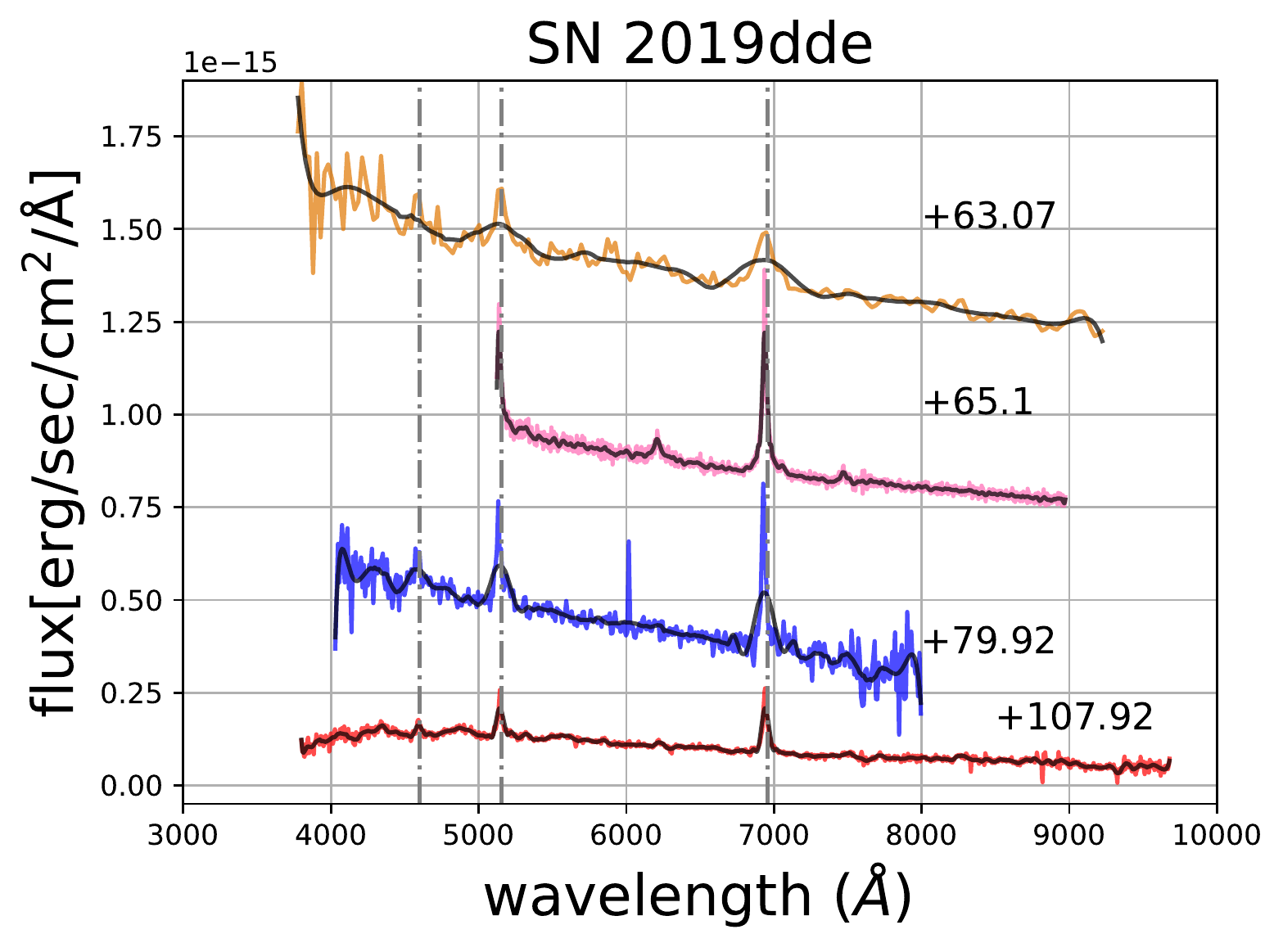}
\includegraphics[scale=.35]{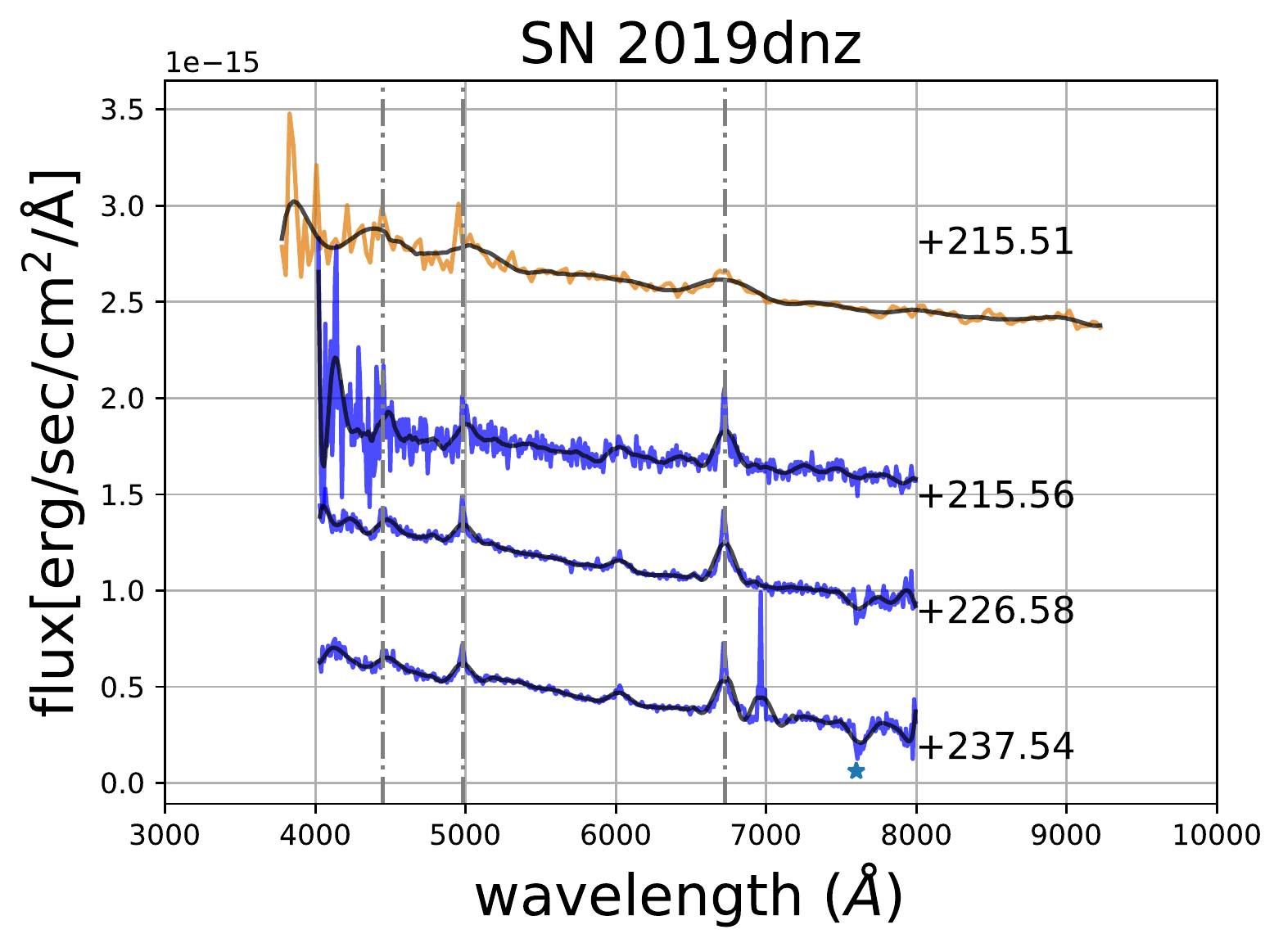}
\caption{Optical spectra of all Type IIn SNe studied for this article. The dashed vertical lines show the Balmer series. The blue stars indicate telluric absorption.}
\label{fig:spectra}
\end{center}
\end{figure*}

All spectra are shown in Figure~\ref{fig:spectra} and are available from WISeREP. In the following, we summarize the reduction procedures applied for each spectrum. All spectroscopic observations were %was reduced using {\color{red} [?]}. It 
calibrated in the following way: since we have contemporaneous P48 $r$-band data, all spectra were scaled so that their synthetic photometry matches the P48 $r$-band value. 

%\newline
%{\color{red} For the following, I think we need to rethink the way we present the technical info about each object. Not sure one paragraph per object only to give the technicalities is necessary, what do you think? On the other hand I don't want to put this info in the anlysis section because it is not analysis really... maybe a table? (but for some object there is quite a lot of info..)}

The Spectral Energy Distribution Machine (SEDm, \citealt{Benami2012,Blagorodnova2018}) spectra were automatically reduced by the IFU data reduction pipeline \citep{Rigault2019}.

The SPRAT spectra were processed by a modification of the pipeline for FrodoSpec \citep{Barnsley2012}. 

The spectra taken with the Andalucia Faint Object Spectrograph and Camera (ALFOSC), mounted on the 2.56-meter Nordic Optical Telescope (NOT), were reduced following standard IRAF\footnote{{IRAF} is distributed by the National Optical Astronomy Observatories, which are operated by the Association of Universities for Research in Astronomy, Inc., under cooperative agreement with the National Science Foundation.} procedures.

The spectra taken with the Auxiliary-port CAMera (ACAM), mounted on the 4.2-m William Herschel Telescope (WHT), were processed following standard IRAF procedures. 

The data from the Double Beam Spectrograph (DBSP) on the Palomar 200-inch (P200) telescope were reduced following standard IRAF procedures of long slit spectroscopy. The two-dimensional (2D) images were first bias subtracted and flatfield-corrected, then the 1D spectral spectra were extracted, wavelength calibrated with comparison lamps, and flux calibrated using observations of spectrophotometric standard stars observed during the same night and at approximately similar airmasses to the SN.

The spectra taken with the SuperNova Integral Field Spectrograph (SNIFS; Aldering et al. 2002; Lantz et al. 2004) were obtain from TNS with kind permission from Anna V Payne and Michael A. Tucker.

Data taken with the FLOYDS spectrograph mounted on the 2m Faulkes Telescope North, Hawaii, USA through the observing program  TAU2019A-008. A 1\farcs{2} slit was placed on the target. The spectrum was extracted and calibrated following standard procedures using the FLOYDS data reduction pipeline\footnote{\url{https://github.com/svalenti/FLOYDS\_pipeline}} \citep{Valenti_etal_2014}. 

Data from the Dual Imaging Spectrograph (DIS) mounted on the 3.5 m Astrophysics Research Consortium (ARC) telescope at the Apache Point Observatory were reduced using standard procedures and calibrated to a standard star obtained the same night using the PyDIS package \citep{pydis};

Data taken with the the Keck Low-Resolution Imaging Spectrometer (LRIS) \citep{Keck}. The data was reduced with the LRIS automated reduction pipeline\footnote{\url{http://www.astro.caltech.edu/∼dperley/programs/lpipe.html}}\citep{Perley2019}.

\begin{deluxetable}{lll}
\tablecolumns{3}
\tablecaption{Summary of spectroscopic observations}
\tablewidth{0pt}
\tablehead{\colhead{Object}&\colhead{Date}&\colhead{Facility}}%&\colhead{Reference}}
\startdata
SN\,2018bwr &2018 Jun 02&P60 + SEDM\\
&2018 Jun 10&LT + SPRAT\\
& 2019 Jan 19 & NOT + ALFOSC\\%&&-\\
SN\,2018lpu &2018 Jul 17&P200 + DBSP [1]\\
SN\,2018fdt &2018 Aug 12&P60 + SEDM\\%&-\\
&2018 Aug 19&P200 + DBSP \\%&\cite{Oke1982}\\
SN\,2018gwa &2018 Oct 06&P60 + SEDM\\%&&-\\
&2018 Oct 10&P200 + DBSP\\%&\cite{Oke1982}\\
SN\,2018kag &2018 Dec 24&P60 + SEDM\\%&-\\
&2019 Jan 19&NOT + ALFOSC\\%&&-\\
& 2019 Feb 20 & WHT + ACAM \\
SN\,2019qt & 2019 Jan 13 & UH88 + SNIFS * \\
& 2019 Feb 04 & NOT + ALFOSC\\
SN\,2018lnb & 2019 Jan 29 & LT + SPRAT \\
& 2019 Feb 04 & LCOGT 2m + FLOYDS  \\
& 2019 Feb 04 & LT + SPRAT \\
& 2019 Feb 12 & P200 + DBSP \\
& 2019 Feb 12 & P200 + DBSP \\
& 2019 Mar 15 & NOT + ALFOSC \\
& 2019 Mar 15 & P60 + SEDM \\
SN 2019cac & 2019 Mar 14 & P60 + SEDM \\
& 2019 Apr 07 & NOT + ALFOSC \\
SN\,2019cmy & 2019 Mar 29 & P60 + SEDM\\
&2019 Mar 30 & ARC + DIS \\
&2019 Mar 30 & P60 + SEDM \\
&2019 Mar 30 & P60 + SEDM \\
&2019 Mar 31 & P60 + SEDM \\
&2019 Apr 03 & Keck1 + LRIS \\
&2019 Apr 07 & ARC + DIS \\
SN\,2019ctt & 2019 Apr 06 & UH88 + SNIFS * \\
& 2019 Apr 22 & P60+SEDM \\
& 2019 Apr 24 & P200+DBSP \\
SN\,2019dde & 2019 Apr 14 & P60 + SEDM \\
& 2019 Apr 16 & SOAR + Goodman * \\
& 2019 May 01 & LT + SPRAT \\
SN\,2019dnz & 2019 Apr 19 & P60 + SEDM  \\
& 2019 Apr 19 & LT + SPRAT \\
& 2019 Apr 30 & LT + SPRAT \\
& 2019 May 11 & LT + SPRAT
\enddata
\tablecomments{The spectra marked with a star were obtained from the TNS and kindly made available to us by Anna V Payne, Michael A. Tucker (SCAT) and Dr. Regis Cartier. [1] The 600/4000 grism and 316/7500 grating were used for the blue and red cameras, respectively, with the D55 dichroic.} %{\color{red}[add error \label{tab:format}}
\label{table:obs}
\end{deluxetable}

\section{Analysis}

\subsection{Epoch of zero flux}\label{sec:t0}
In order to derive the extrapolated epoch of zero flux of all the events, we used the {\tt Photomanip}\footnote{https://github.com/maayane/PhotoManip} package (released in the Appendix of this paper) to fit the $r$-band flux during the rise time (or the $g$-band flux light curve, when early $r$-band data points are not available) with an exponential function of the form
\begin{equation}\label{eq:bolometric luminosity exp}
f=f_{\rm{max}}\{ 1-\exp{[(t_{0}-t)/t_c]} \}\;,
\end{equation}
and a power-law of the form
\begin{equation}\label{eq:bolometric luminosity power}
f=a(t-t_{0})^n \;,
\end{equation}

(where $t_{0}$ is the extrapolated time of zero flux, $f_{\rm{max}}$ is the maximum flux, $t_{c}$ is the characteristic rise time of the $r$-band light curve). In each case, we chose the function giving the best fit (i.e. lowest $\chi^2/dof$), which allowed us to estimate the epochs at which the extrapolated light curves are reaching zero, which are used throughout this paper as the reference time $t_{0}$, and are  summarized in Table~\ref{table:t0}. For each SN in our sample, the table shows the band in which the fit was performed ($g$ or $r$, depending on how constraining the data are), the prior on $t_0$ is taken to be a time-interval from $\sim1$\,day before the most recent pre-explosion upper limit and the first detection. Table~\ref{table:t0} also shows the $1\sigma$ confidence interval on $t_{0}$. The typical uncertainty on $t_{0}$ is of order $1$ to a few days, with the exception of SN\,2019cac (where no previous non-detection exists and for which we applied a broad conservative prior on $t_0$), for which it is higher than $20$ days.

\begin{deluxetable*}{lcccccc}
\tablecolumns{7}
\tablecaption{Reference times fitting results}
\tablewidth{0pt}
\tablehead{\colhead{IAU Name}&\colhead{ZTF Name}&\colhead{model}&\colhead{band}&\colhead{most recent upper limit}&\colhead{$t_{0}$}&\colhead{confidence interval}\\
\colhead{} & \colhead{} & \colhead{} & \colhead{} & \colhead{($MJD$)} & \colhead{($MJD$)} &\colhead{($MJD$)}}
\startdata
%SN\,2018lpu & ZTF18abgrlpv&power law&$g$&$2458306.847$&$2458306.846$&[$2458306.845$,$2458307.330$]\\
%SN\,2018fdt & ZTF18abltfho&exponent&$r$&$2458334.665$&$2458336.335$&[$2458335.717$,$2458336.595$]\\
%SN\,2018gwa & ZTF18abxbhov&exponent&$g$&$2458374.969$&$2458376.540$&[$2458374.708, 2458376.542$]\\ \hline 
%SN\,2018bwr & ZTF18aavskep&exponent&$r$&$2458257.523$&$2458257.527$&[$2458257.360$, $2458257.627$]\\
%SN\,2018kag & ZTF18acwzyor&power law&$g$&$2458464.965$&$2458467.581$&[$2458466.039$, $2458467.870$]\\
%SN\,2019qt & ZTF19aadgimr&exponent&$g$&$2458488.008$&$2458491.728$&[$2458491.627, 2458491.796$]\\
%SN\,2018lnb & ZTF19aaadwfi&power law&$g$&$2458467.972$&$2458470.331$&[$2458468.259,2.458472.329$]\\
%SN\,2019cac & ZTF19aaksxgp&power law&$g$&$2458521.778$&$2458521.937$&[$2458503.976$,$2458526.771$]\\
%SN\,2019cmy & ZTF19aanpcep &exponent&$g$&$2458567.983$&$2458568.502$&[$2458568.324$,$ 2458568.342$] \\
%SN\,2019ctt & ZTF19aanfqug &exponent&$r$&$2458541.796$&$2458550.024$&[$2458546.666$,$2458551.83$]\\
%SN\,2019dde & ZTF19aaozsuh &power law&$r$&$2458573.902$&$2458582.441$&[$2458580.050$,$2458582.741$]\\
%SN\,2019dnz & ZTF19aaqasrq &exponent&$r$&$2458581.995$&$2458583.425$&[$2458582.851$,$2458583.682$]
SN\,2018lpu & ZTF18abgrlpv&power law&$g$&$58306.35$&$58306.35$&[$58306.35$,$58306.83$]\\
SN\,2018fdt & ZTF18abltfho&exponent&$r$&$58334.17$&$58335.83$&[$58335.22$,$58336.10$]\\
SN\,2018gwa & ZTF18abxbhov&exponent&$g$&$58374.47$&$58376.04$&[$58374.21$,$58376.04$]\\ \hline
SN\,2018bwr & ZTF18aavskep&exponent&$r$&$58257.02$&$58257.03$&[$58256.86$,$58257.13$]\\
SN\,2018kag & ZTF18acwzyor&power law&$g$&$58464.46$&$58467.08$&[$58465.54$,$58467.37$]\\
SN\,2019qt & ZTF19aadgimr&exponent&$g$&$58487.51$&$58491.23$&[$58491.13$,$58491.30$]\\
SN\,2018lnb & ZTF19aaadwfi&power law&$g$&$58467.47$&$58469.83$&[$58467.76$,$58471.83$]\\
SN\,2019cac & ZTF19aaksxgp&power law&$g$&$58521.28$&$58521.44$&[$58503.48$,$58526.27$]\\
SN\,2019cmy & ZTF19aanpcep &exponent&$g$&$58567.48$&$58568.00$&[$58567.82$,$58567.84$]\\
SN\,2019ctt & ZTF19aanfqug &exponent&$r$&$58541.30$&$58549.52$&[$58546.17$,$58551.33$]\\
SN\,2019dde & ZTF19aaozsuh &power law&$r$&$58573.40$&$58581.94$&[$58579.55$,$58582.24$]\\
SN\,2019dnz & ZTF19aaqasrq &exponent&$r$&$58581.50$&$58582.92$&[$58582.35$,$58583.18$]\
\enddata
\tablecomments{The "model" column specifies whether a power law (Equation~\ref{eq:bolometric luminosity power}) ore a concave exponent (Equation~\ref{eq:bolometric luminosity exp}) gives the best fit. The "band" column specifies the band ($g$ or $r$) used for the fit, and was chosen according to the amount of data available in each band. We then report the most recent non detection, which we use as the lower limit of our prior on $t_0$ (we use the most recent detection as the upper limit). For SN\,2019cac, no previous non-detection exists, and so our prior interval is a time interval of $100$ days before the first detection. The "$t_{0}$" column is the best fit time at which the flux reaches zero - the time used as an estimate of the explosion epoch. The confidence interval, shown in the last column, is defined here as the tightest intervals containing $68\%$ of the probability and including our best-fit $t_0$ value.} %{\color{red}[add error \label{tab:format}}
\label{table:t0}
\end{deluxetable*}

\subsection{Blackbody temperature, radius and bolometric luminosity}
\begin{figure}\label{fig:evo}
\begin{center}
\includegraphics[scale=.43]{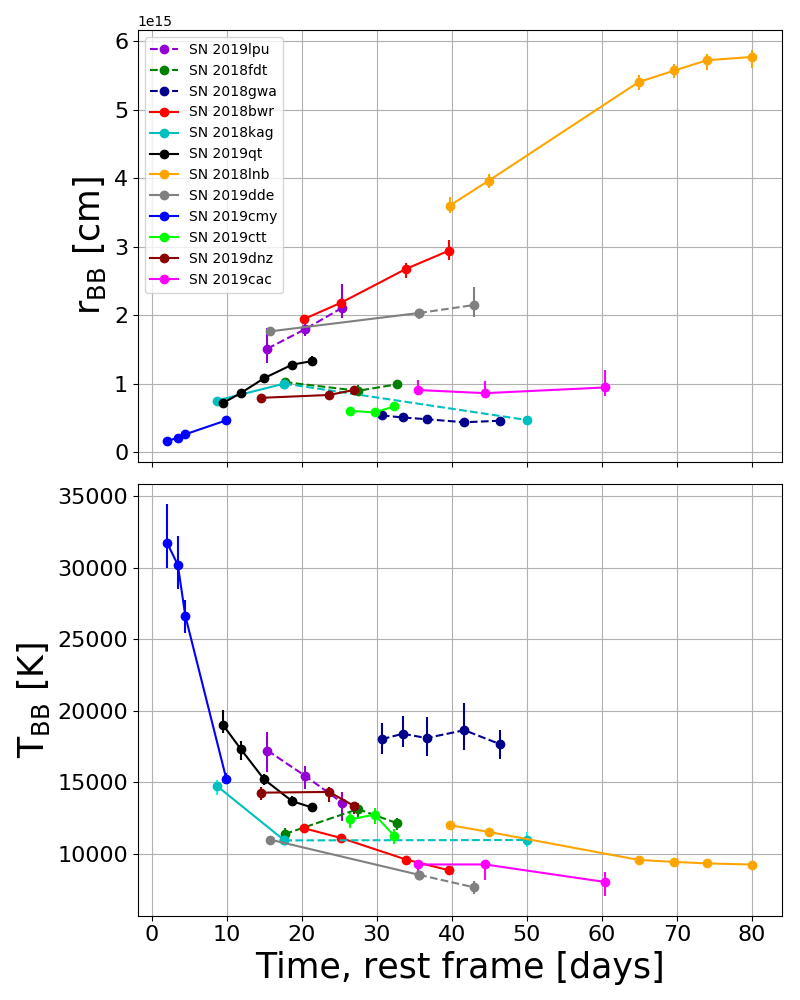}
\caption{The evolution in time of: (1) the radius (upper panel), (2) the temperature (lower panel) of a blackbody with the same radiation as each of the twelve SNe in our sample. The points were obtained by fitting a blackbody spectrum to the observed photometry, after interpolating the various data sets to obtain data coverage of coinciding epochs. The errors were obtained with Monte Carlo Markov chain simulations. The dashed lines correspond to objects for which no late spectra was obtained in order to confirm that the CSM is optically thick. They should be taken cautiously.} %The stars indicate the values derived by fitting a blackbody to the spectroscopic data. The dashed line in the top panel shows the best linear fit to the rising radius phase: a linear function with a slope of $\approx8000$\,km\,s$^{-1}$. At late times, the blackbody model for the spectral energy distribution may not be valid anymore (see e.g. right panel in figure~\ref{fig:spectra}): these points are shown in grey to emphasise that they are less reliable and should be taken cautiously.} 
\label{fig:evo_param_TR}
%\end{center}
\end{center}
\end{figure}

Taking advantage of the multiple-band photometry coverage, we used the {\tt PhotoFit}\footnote{https://github.com/maayane/PhotoFit} tool \citep{Soumagnac2019_2018fif} to derive the temperature and radius of the blackbody that best fits the photometric data at each epoch. The derived best-fit temperatures $T_{BB}$ and radii $r_{BB}$ are shown in Figure ~\ref{fig:evo_param_TR}. We observe that seven  objects of our sample exhibit a fast increase of the blackbody radius, a result in contrast with most previous observations. Indeed, many previously studied SNe IIn showed a constant blackbody radius (e.g., SN2010jl; \citealt{Ofek2014}), consistent with the continuum photosphere being located in the unshocked optically thick CSM. In some cases a blackbody radius stalling after a short increase (e.g., 2005kj, 2006bo, 2008fq, 2006qq, \citealt{Taddia2013}; 2006tf, \citealt{Smith2008}) or even a shrinking blackbody radius (e.g., SN2005ip; SN2006jd, \citealt{Taddia2013}) were observed. Such observations were explained by the possible presence of clumps in the CSM that may expose underlying layers \citep{Smith2008}. PTF\,12glz was not the only case were a fast increase of the blackbody radius was observed: three of the SNe IIn observed - in the UV - by \cite{DelaRosa2016} showed blackbody radii growing at comparable rates. This could be due to the fact that UV observations provide a better handle on the blackbody spectrum shape than visible light alone, suggesting that a fast increase of the blackbody radius of SNe IIn may be more common than suggested by visible-light surveys of these objects. 

We further discuss and exploit the $r_{BB}$ measurement in \S~\ref{sec:aspherical}. 
%{\color{blue} in section 4 we talk of 6 objects with rising rbb, not 7. This is because the analysis of section 4 is not applicable to one of them. But we should probably explain the discrepency somewhere}.

\begin{figure}\label{fig:evo}
\begin{center}
\includegraphics[scale=0.55]{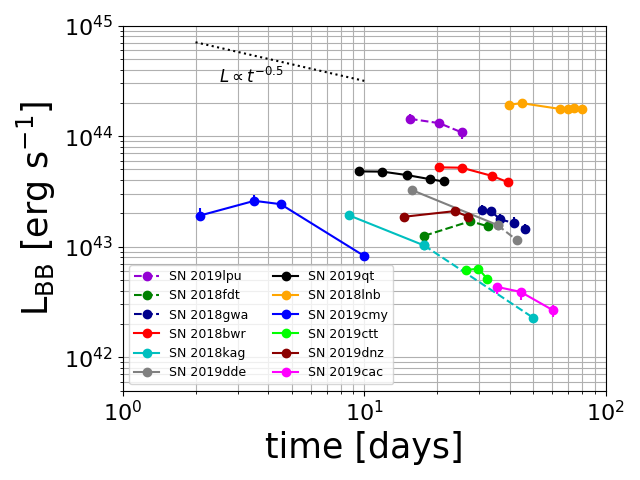}
\caption{The evolution in time of the bolometric luminosity of a blackbody with the same radiation as each of the twelve SNe in our sample. The dashed lines correspond to objects for which no late spectra was obtained in order to confirm that the CSM is optically thick. They should be taken cautiously. The dotted line shows the $t^{-0.5}$ slope (see e.g. \citealt{Ofek2014}).} 
\label{fig:evo_param_L}
%\end{center}
\end{center}
\end{figure}

Based on the measurement of $r_{BB}$ and $T_{BB}$, we were able to derive the luminosity $L_{\rm BB}=4\pi r_{\rm BB}^2\sigma T_{\rm BB}^4 $ of the blackbody fits, shown in Figure~\ref{fig:evo_param_L}.

\subsection{Spectroscopy}\label{sec:analysis_spectroscopy}

In this section, we only report the spectroscopic information that allow us to assess which photometric data are usable for our analysis of the CSM geometry. Indeed, the asphericity criterion proposed by \cite{Soumagnac2019} is only applicable at times when the CSM surrounding the explosion is optically thick. To verify this, we require that the spectrum will be dominated by a blackbody continuum with no high velocity ($\gtrsim2000\,\rm km\,s^{-1}$) absorption and emission lines.

We can only include in our analysis multiple-band photometry that was collected before, or close to, the observation of a spectrum showing no evidence for high-velocity material. Unfortunately, we were unable to secure such spectroscopy for the SNe IIn  SN\,2018lpu,  SN\,2018fdt and SN\,2018gwa, for which no spectra were taken after or close to the last $Swift$ data point.

\subsubsection{SN\,2018bwr}
 
The first two spectra show H$_\alpha$, H$_\beta$ and H$_\gamma$ emission lines. In the last spectrum, we see prominent broad \ion{Ca}{2} emission, blended with the \ion{O}{1} $\lambda\,8446$\,\AA\; feature. The numerous Fe lines are blended, exhibiting a pseudo-continuum around $\sim$5500\,\AA. Such a pseudo continuum is also seen e.g. in PTF\,12glz \citep{Soumagnac2019} and in SN\,2005\,cl \citep{Kiewe2012}. We conclude from this that the spectra are dominated by interaction out to late times, and we can use all of the UV photometry for our analysis.

\subsubsection{SN\,2018kag}

The first spectrum shows a blue continuum with Balmer emissions lines. The Balmer lines remain discernible at +30.3 d and the continuum becomes flat. At +63.10 d, higher velocity absorption and emission lines have appeared in the spectrum, hinting that the CSM may not be optically thick anymore. As a result, only the UV photometry taken between the first two spectra is usable for our analysis of the CSM geometry.

\subsubsection{SN\,2019qt}

Distinct narrow H$_\alpha$ and H$_\beta$ emission lines are visible in both spectra. H$_\gamma$ emission is also visible, especially in the earlier spectrum. Since all the UV photometry was taken between the epochs of these two spectra, all of it is usable for our analysis.

\subsubsection{SN\,2018lnb}
Narrow Balmer emission are visible in all spectra except for the first two spectra, in which the H$\alpha$ component falls outside the spectral range of SPRAT/LT, and the SEDm/P60 spectrum which has low signal-to-noise. All of the UV photometric data is usable for our analysis.

\subsubsection{SN\,2019cac}

In spite of the low resolution of the first spectrum, H$_\alpha$ emission is visible at +31.9 d.
Strong emission lines of H$_\alpha$, H$_\beta$ and H$_\gamma$ can be observed at +55.9 d. Although the last UV data point was taken after the second spectrum, we consider their epochs to be close enough so that all of the UV data can be used for our analysis. %however, in this same spectrum, emission line of He (4472) are even stronger than the one of H$_\alpha$.

\subsubsection{SN\,2019cmy}

%The narrow Balmer emission lines, that define the type IIn class, are the signature of an external physical phenomenon highly dependent on the surrounding environment, rather than of any intrinsic property of the explosion. In the case of flash-spectroscopy events, these lines only persist for days, whereas in the case of SNe IIn they may still be visible in the spectrum for months or years. %and in intermediate cases they may persists for weeks (e.g., SN\,1998s, \citealt{Li1998, Fassia2000, Fassia2001}; SN\,2005gl, \citealt{Gal-Yam2007}; SN\,2010mc, \citealt{Ofek2013}). 
%The type IIn class is not a well-defined category of objects, and in particular, t
The limit between flash-spectroscopy events and Type IIn SNe can be blurry, when the Balmer lines persist for weeks or a few months.%as many SNe show the characteristic narrow Balmer lines in their spectra, sometime during their evolution.

%SN\,2019cmy could be defined as a ``long-lived flasher'' rather than a proper SN IIn.
In the case of SN\,2019cmy, prominent narrow Balmer emissions lines are visible at +4.9 d, with the characteristic broad wings of the H$\alpha$ line, interpreted as the signature of electron scattering, clearly visible. Strong high-ionization emission lines of \ion{He}{2} $\lambda\,4686$\,\AA\; only persists at +4.9 days. An excess on the blue side of the \ion{He}{2} $\lambda\,4686$\,\AA\; coincides with the high-ionized \ion{C}{3} $\lambda\,4650$\,\AA. However, by  +8.5 d, the \ion{C}{3} $\lambda\,4650$\,\AA\; and \ion{He}{2} $\lambda\,4686$\,\AA\; lines have completely disappeared, consistent with flash-ionized emissions. The Balmer lines decrease in strength with time: the H$_\gamma$ $\lambda\,4341$\,\AA\; and H$_\delta$ $\lambda\,4102$\,\AA\; are marginally detected at +8.5 d and have disappeared by day +12.4. A spectrum taken two months after first light (and not shown in this paper) exhibits the features of a ``normal'' Type II SN, without any particular signature of CSM interaction.

Our geometrical analysis, which probes the shape of the CSM rather than its amount or the physical ways by which it was ejected, should still hold. All the UV photometry is usable for our analysis.

\subsubsection{SN\,2019ctt}

Narrow Balmer lines (H$_\alpha$, H$_\beta$, H$_\gamma$) are visible in all three spectra. The H$_\delta$ line is also visible in the higher resolution spectrum at +32.4 d. All the UV photometry is usable for our analysis.

\subsubsection{SN\,2019dde}
The first spectrum, taken at +63.07 d with the SEDm/P60 shows narrow Balmer lines (H$_\alpha$, H$_\beta$, H$_\gamma$, H$_\delta$, H$_\epsilon$). The three later spectra at +65.10 d, +79.92 d and +107.92 d show narrow H$_\alpha$ and H$_\beta$ emission lines, 

In the last spectrum, a narrow He $\lambda\,5876$\,\AA\; emission line is visible. Although the Balmer series is strongly dominated by narrow emission, the broad absorption at 5000-10000 $\rm km\,s^{-1}$ suggests that the ejecta have become visible, and the CSM is not completely optically thick anymore.

To account for this, we only use the first two UV epochs for our analysis.

\subsubsection{SN\,2019dnz}
 Narrow Balmer lines (H$_\alpha$, H$_\beta$, H$_\gamma$) are visible in all three spectra. In addition, H$_\delta$, H$_\epsilon$ emission lines can be seen in the last spectrum. All the UV photometry can be used for our analysis.

\subsubsection{Events with missing final spectra}\label{sec:missing}

For three objects in our sample, we were unable to collect a spectrum showing no evidence for high-velocity material close to or after the last UV photometry epoch. For SN\,2018\,lpu, one spectrum was taken, where strong and narrow Balmer lines can be seen. Other interesting features include narrow emission of \ion{He}{2} ($\lambda\,3203$\,\AA, $\lambda\,4686$\,\AA), [\ion{O}{2}]$\lambda\,3727$\,\AA, and [\ion{O}{3}] $\lambda\,5007$\,\AA.  For both SN\,2018fdt and SN\,2018gwa, two spectra were obtained before any $Swift$ photometry was taken. Both show prominent narrow Balmer emissions lines.

\section{Fraction of SNe IIn showing evidence for aspherical CSM}\label{sec:aspherical}

In all that follows, we assume that the criterion from \cite{Soumagnac2019}, i.e. a fast increase of the blackbody effective radius (if observed at times when the CSM surrounding the explosion is still optically thick), is an indication for asphericity. We note that asphericity could manifest in other ways, and that this approach does not allow us to exclude or constrain more complicated geometries.

\subsection{Application of the asphericity criterion from \cite{Soumagnac2019}}\label{sec:criterion}

\begin{figure}
\begin{center}
\includegraphics[scale=.43]{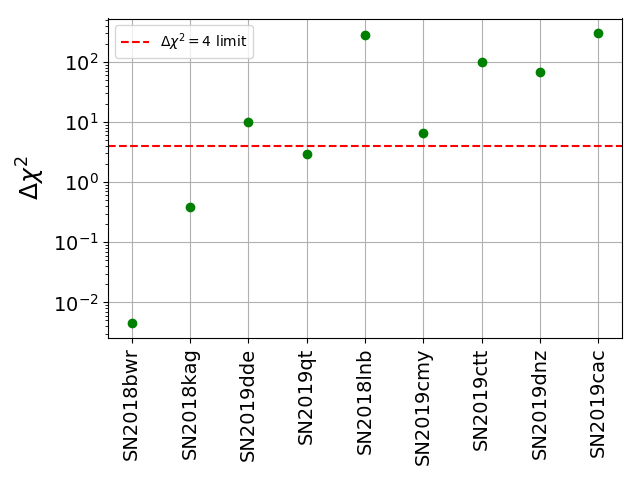}
\caption{Result of the likelihood-ratio test (or chi-square difference test), when modeling the evolution of $r_{\rm BB}$ with a power law and with a flat function. The red dashed line shows the $\Delta \chi^2=4$ (i.e. $2\sigma$) limit for one degree of freedom difference: objects with a $\Delta \chi^2$ limit above this line are better modeled by a non-zero power law (and hence show evidence for aspherical CSM), whereas objects below this line are better modeled by a flat line (i.e. show no evidence for aspherical CSM). Applying the criterion from \cite{Soumagnac2019}, six out of nine SNe IIn in our sample show evidence for aspherical CSM. The SNe are ordered by their maximum measured bolometric luminosity (left to right).} 
\label{fig:deltachi2}
%\end{center}
\end{center}
\end{figure}

Assessing whether the blackbody radius $r_{\rm BB}$, shown in Figure~\ref{fig:evo_param_TR}, is growing or not, is a hypothesis testing problem, i.e. we need to select between two models the one that best explains the data. Our model is a power law function of the form $R=R_0\left(\frac{t}{t_0}\right)^n$, where the null hypothesis is that $n=0$ and the alternative hypothesis is that $n\neq 0$. %and our second model is a flat function of the form $R=R_0$ (i.e. $n=0$). 
Since these models are nested, we can apply a likelihood-ratio test (or chi-square difference test) to discriminate between them.  In Figure~\ref{fig:deltachi2}, we show the $\chi^2$ difference between the two models derived for all objects. For six out of nine objects, $\Delta \chi^2>4$ i.e. the chi-square difference indicates that the increasing radius is more likely than the constant radius at a $2\sigma$ level. Therefore $66\%$ of the SNe in our sample (taking into account only the SNe to which our analysis is applicable) show evidence for aspherical CSM.

\subsection{Do brighter SNe IIn have more aspherical CSM? correction for potential selection effects}\label{sec:corr}

%\subsection{Correction for the non-uniform volume distribution of the SNe in our sample}\label{sec:corr}

\begin{figure}
\begin{center}
\includegraphics[scale=.43]{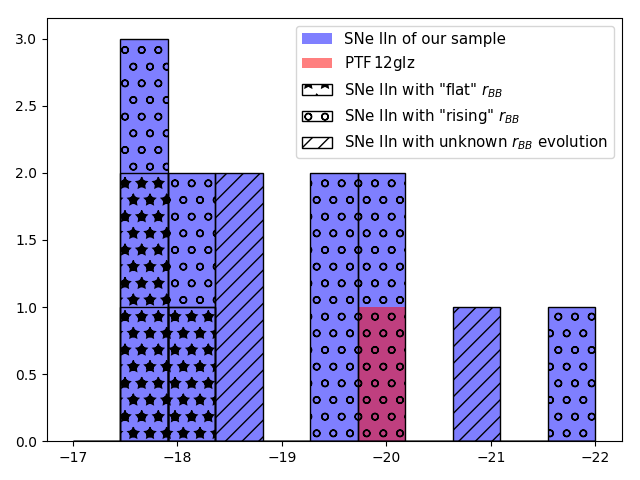}
\caption{Absolute magnitude of the twelve SNe IIn of our sample and PTF\,12glz. The blue histograms correspond to the entire sample and the red square corresponds to PTF\,12glz. The star-patterned histograms correspond to the SNe IIn whose radius is better modeled by a flat function than by a power law (i.e. showing no evidence for aspherical CSM). These objects are at the faint end of the distribution, an effect we need to correct for in the calculation of their probability to occur (see \S~\ref{sec:corr}). The circle-patterned histograms correspond to the SNe IIn whose radius is better modeled by a power law (i.e. showing evidence for aspherical CSM). The line-patterned histograms correspond to the SNe IIn discussed in \S~\ref{sec:missing}, i.e. for which no late spectrum was collected and to which our analysis of the CSM geometry does not apply. } 
\label{fig:stat1}
\end{center}
\end{figure}

\begin{figure*}
\begin{center}
\includegraphics[scale=.43]{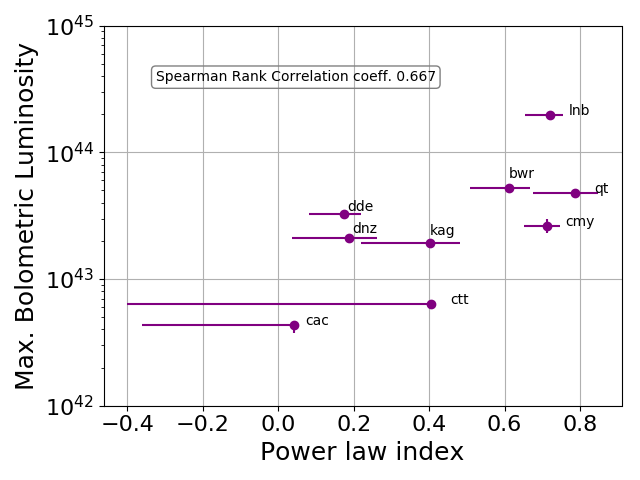}
\includegraphics[scale=.43]{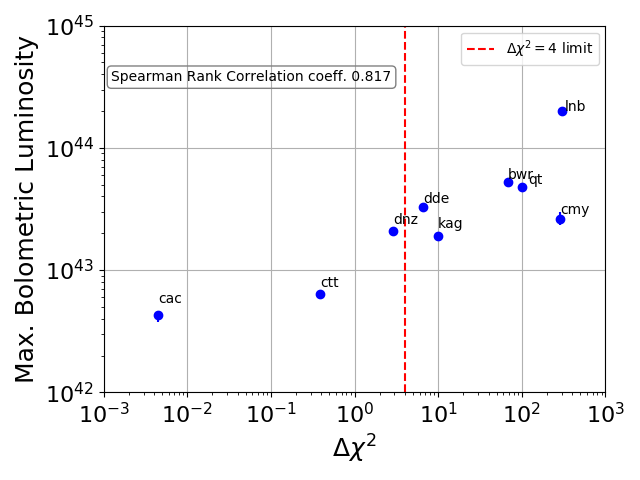}
\caption{Lower limit on the peak bolometric luminosity as a function of: (left panel) the index of the power law that best fits the blackbody radius and (right panel) the $\chi^2$ difference between the two models derived for all objects (see \S~\ref{sec:criterion}). Both quantities are assumed to be related to the asphericity of the CSM, and appear to be correlated with the peak bolometric luminosity.} 
\label{fig:ind_maxbolo}
%\end{center}
\end{center}
\end{figure*}

In Figure~\ref{fig:stat1}, we show the distribution of absolute magnitudes of the SNe IIn in our sample. The overall distribution (in blue) is comparable to previously published absolute luminosity distribution for SNe IIn (see e.g. Figure 17 in \citealt{Richardson2014}). However, the SNe showing no evidence for a rising $r_{\rm BB}$ are on the faint end of the distribution. This trend is also visible in Figure~\ref{fig:ind_maxbolo}, where we show the lower limit on the bolometric luminosity of all the SNe IIn in our sample (also reported in Table~\ref{tab:bolochi2}), as a function of (1) the index of the power law that best fits $r_{BB}$ and (2) the $\chi^2$ difference between the two models derived for all objects (also reported in Table~\ref{tab:bolochi2}; see \S~\ref{sec:criterion}). 

The objects of our sample which are intrinsically brighter appear to show evidence for an increasing blackbody radius -- which we interpret as an indication for aspherical CSM -- whereas fainter objects tend not to show such feature. The Spearman rank correlation between the power law index and the lower limit on the bolometric luminosity is $0.67$, and $0.82$ between the lower limit on the bolometric luminosity and $\Delta \chi^2$. The false alarms probabilities are $0.03$ and $0.005$, respectively (the false alarm probability were estimated using bootstrap simulations implemented in \citealt{Ofek_matlab}).

There are several possible explanations to this correlation.
It could be the result of either some selection bias or some physical reasons (or a combination of both). Among the possible physical reasons are the following: (i) More massive, or
alternatively more energetic explosions, tend to occur in aspherical CSM; (ii) A geometrical effect, related to the viewing angle, could also be playing a role. Indeed, if  one thinks about a slab of CSM (for simplicity), the increase of the blackbody radius, which we used in this paper as a criterion for asphericity, is most patent when the explosion is observed perpendicularly to the long axis of the slab. The brighter events of our sample could happen to be observed from this direction, while the fainter events could be observed from the short axis direction, preventing us from detecting the asphericity of their CSM using our criterion. We plan to explore this effect in future work.

The observed correlation could also be due to some selection effects. If we assume that both classes of SNe IIn obey the same luminosity - and volume - distribution, the SNe showing no evidence for a rising $r_{\rm BB}$ appear to be under-represented in our sample, a fact that needs to be corrected for in the final probability calculation. (A full relative rate calculation, taking into account a broader variety of selection effects, e.g. due to the cadence, the varying limiting magnitude of each image or the extinction at the location of the SN, is beyond the scope of this paper). Here, we simply estimate the relative probability $p_i$ of finding the $i^{th}$ SN IIn of our sample (SN$_i$) as

%\begin{equation}
%p_i=\dfrac{\frac{1}{D_{max,i}^3}}{\sum\limits_{j=1}^{9}\frac{1}{D_{max,j}^3}}\;,
%\end{equation}
\begin{equation}
p_i=\dfrac{\frac{1}{V_{max,i}}}{\sum\limits_{j=1}^{9}\frac{1}{V_{max,j}}}\;,
\end{equation}
where $V_{max,i}$ is the maximum volume to which SN$_i$ can be observed, under the assumption of a constant limiting magnitude for the survey in the $r$, $m_{\rm lim}=20.5$.
%and is given by
%\begin{equation}
%D_{max,i}=10^{\frac{m_{\rm lim}-M_i}{5}-5}\;[\rm Mpsec]\;,
%\end{equation}
%where $m_{\rm lim}=20.5$ is the limiting magnitude for the survey in the $r$ band and $M_i$ is the absolute magnitude of SN$_i$.
In case both classes of objects obey the same luminosity distribution, the corrected fraction of SNe IIn exhibiting a rising $r_{\rm BB}$ is $35\%$. 

To conclude, depending on the assumption we make on the luminosity distribution of both classes of objects, the fraction of SNe IIn showing an increasing radius, deduced from our sample, could be $35\%$, or as high as $66\%$. As this is a sufficient but not necessary condition for the surrounding CSM to be aspherical, these numbers are a lower limit on the fraction of SNe IIn exploding in aspherical CSM.

%Untill further progress is done in determining the relative luminosity distribution of both classes of objects, we conclude that the lower limit on the fraction of SNe IIn showing a rising $r_{\rm BB}$ is between $35\%$ and 
%As this is a sufficient but not necessary condition for the surrounding CSM to be aspherical, $35\%$ is a lower limit on the fraction of SNe IIn exploding in aspherical CSM.

\begin{comment}
\begin{deluxetable*}{lcccccc}
\tablecolumns{5}
\tablecaption{}
\tablewidth{0pt}
\tablehead{\colhead{IAU Name}&\colhead{lower limit on L}&\colhead{$\Delta\chi^2$}&\colhead{power law index}&\colhead{confidence interval}}\\
\startdata
SN2018bwr&$5.22\times10^{43}$&$68.6$&$0.61$&$[0.51,0.67]$\\
SN2018kag&$1.92\times10^{43}$&$9.8$&$0.4$&$[0.22,0.48]$\\
SN2019dde&$3.26\times10^{43}$&$6.5$&$0.17$&$[0.08,0.22]$\\
SN2019qt&$4.80\times10^{43}$&$99.3$&$0.79$&$[0.68,0.85]$\\
SN2018lnb&$19.87\times10^{44}$&$302.0$&$0.72$&$[0.66,0.75]$\\
SN2019cmy&$2.60\times10^{43}$&$284.6$&$0.71$&$[0.65,0.75]$\\
SN2019ctt&$6.32\times10^{42}$&$0.4$&$0.41$&$[-0.4,0.41]$\\
SN2019dnz&$2.10\times10^{43}$&$2.9$&$0.19$&$[0.04,0.26]$\\
SN2019cac&$4.33\times10^{42}$&$0.0$&$0.04$&$[-0.36,0.04]$
\enddata
\tablecomments{} %{\color{red}[add error \label{tab:format}}
\label{table:t0}
\end{deluxetable*}
\end{comment}

\begin{deluxetable}{lcc}
\tablecolumns{3}
\tablecaption{Bolometric luminosity and asphericity of the CSM}
\tablewidth{0pt}
\tablehead{\colhead{IAU Name}&\colhead{lower limit on peak $L_{\rm BB}$}&\colhead{$\Delta\chi^2$}\\
\colhead{} & \colhead{$[\rm{erg/s}]$}}
\startdata
SN2018bwr&$5.22\times10^{43}$&$68.6$\\
SN2018kag&$1.92\times10^{43}$&$9.8$\\
SN2019dde&$3.26\times10^{43}$&$6.5$\\
SN2019qt&$4.80\times10^{43}$&$99.3$\\
SN2018lnb&$19.87\times10^{44}$&$302.0$\\
SN2019cmy&$2.60\times10^{43}$&$284.6$\\
SN2019ctt&$6.32\times10^{42}$&$0.4$\\
SN2019dnz&$2.10\times10^{43}$&$2.9$\\
SN2019cac&$4.33\times10^{42}$&$0.0$
\enddata
\tablecomments{Lower limit on the peak bolometric luminosity $L_{\rm BB}$ and $\chi^2$ difference between a power law model with $n\neq 0$ and $n=0$.} 
\label{tab:bolochi2}
\end{deluxetable}

\section{Conclusions}\label{sec:discussion}
We presented the first planned Ultra-Violet (UV) survey of the early evolution of type IIn supernovae (SNe IIn).  Our sample consists of 12 SNe IIn discovered and observed with the Zwicky Transient Facility (ZTF) and followed-up in the UV by the {\it Neil Gehrels Swift Observatory}. All SNe were also spectroscopically followed-up: we present and release the spectroscopic data we collected.

The UV observations presented in this paper could help shed light on various aspects of the physical picture governing these events. 
For example, they may be used to better understand the explosion mechanism and the CSM properties (e.g., \citealt{Ofek2013a}), since the collisionless shock propagating in the CSM after the shock breakout \citep{Ofek2010} is predicted to radiate most in the UV and X-rays. 

Observations of SNe IIn at UV wavelengths provide a better handle on the bolometric luminosity, blackbody radius and blackbody temperature than visible-light observations alone. This may be a reason why the fast rising blackbody radius - which we observe for seven objects out of the twelve of our sample - was only observed in the past in works using UV observations of SNe IIn \citep{DelaRosa2016,Soumagnac2019}. This result is in contrast with most previous observations using visible-light observations alone, of either a constant, slowly rising (and then stalling) or even a shrinking blackbody radius. 

Assuming that a rising blackbody radius is an indication for asphericity, we used the UV observations to address the following question: "what fraction of SNe IIn explode in aspherical CSM?". Indeed, although  observations of SNe IIn are usually analyzed within the framework of spherically symmetric models of CSM, resolved images of stars undergoing considerable mass loss as well as well as polarimetry observations, suggest that asphericity is common, and should be taken into account for realistic modeling of these events. Constraining the geometrical distribution of the CSM surrounding the explosion is key to understanding the mass-loss processes occurring before the explosion and the nature of the yet-to-be determined progenitors of SNe IIn. Indeed, the presence of aspherical CSM around the progenitor is hard to explain by a simple wind, and requires to invoke other scenarios, such as episodic emission, rapid stellar rotation, or binarity.

We applied the criterion for asphericity  introduced  by \cite{Soumagnac2019},  stating  that  a  fast  increase  of  the  blackbody effective radius, if observed at times when the CSM surrounding the explosion is still optically thick, may be interpreted as an indication that the CSM is aspherical. We find that two thirds of the SNe in our sample show evidence for aspherical CSM.  We also find that higher luminosity objects tend to show evidence for aspherical CSM. This correlation could be due to physical reasons or to some selection bias. If we assume that both classes of SNe IIn obey the same luminosity - and volume - distribution, the fraction of SNe showing evidence for a rising blackbody radius needs to be corrected. Depending on the assumption we make on the luminosity distribution of both classes of objects, the lower limit deduced from our sample on the fraction of SNe IIn showing evidence for aspherical CSM could be 35\%, or as high as 66\%. This result suggests that asphericity of the CSM surrounding SNe IIn is common -- consistent with data from resolved images of stars undergoing considerable mass loss. It also suggests that asphericity should be taken into account for more realistic modelling of these events.
 
% After  correcting  for selection effects which leads SNe IIn not showing such evidence to be under-represented in our sample, we derive a conservative lower limit of 35\% on the fraction of SNe IIn showing evidence for aspherical CSM.  
 
As future wide-field transient surveys and the {\it ULTRASAT} UV satellite mission \citep{Sagiv2014} are deployed, more UV observations of interracting SNe will be collected, allowing to build upon this survey and to refine the lower limit derived in this paper.

\acknowledgments
M.T.S. thanks Charlotte Ward and Eli Waxman for useful discussions. 

This work is based on observations obtained with the Samuel Oschin Telescope 48-inch and the 60-inch Telescope at the Palomar Observatory as part of the Zwicky Transient Facility project. ZTF is supported by the National Science Foundation under Grant No. AST-1440341 and a collaboration including Caltech, IPAC, the Weizmann Institute for Science, the Oskar Klein Center at Stockholm University, the University of Maryland, the University of Washington, Deutsches Elektronen-Synchrotron and Humboldt University, Los Alamos National Laboratories, the TANGO Consortium of Taiwan, the University of Wisconsin at Milwaukee, and Lawrence Berkeley National Laboratories. Operations are conducted by COO, IPAC, and UW.

We acknowledge the use of public data from the Swift data archive.

SED Machine is based upon work supported by the National Science Foundation under Grant No. 1106171

This paper shows observations made with the Nordic Optical Telescope, operated by the Nordic Optical Telescope Scientific Association at the Observatorio del Roque de los Muchachos, La Palma, Spain, of the Instituto de Astrofisica de Canarias.

Some of the data we present were obtained with ALFOSC, which is provided by the Instituto de Astrofisica de Andalucia (IAA) under a joint agreement with the University of Copenhagen and NOTSA. The Liverpool Telescope, located on the island of La Palma in the Spanish Observatorio del Roque de los Muchachos of the Instituto de Astrofisica de Canarias, is operated by Liverpool John Moores University with financial support from the UK Science and Technology Facilities Council. The ACAM spectroscopy was obtained as part of OPT/2018B/011.

We would like to thank occasional observers on the UW APO ZTF follow-up team, including: Eric Bellm, Zach Golkhou, James Davenport, Daniela Huppenkothen, Dino Bektešević, Gwendolyn Eadie, and Bryce T. Bolin. MLG acknowledges support from the DIRAC Institute in the Department of Astronomy at the University of Washington. The DIRAC Institute is supported through generous gifts from the Charles and Lisa Simonyi Fund for Arts and Sciences, and the Washington Research Foundation.

This work was supported by the GROWTH project funded by the National Science Foundation under Grant No 1545949.

A.G.-Y. is supported by the EU via ERC grant No. 725161, the Quantum Universe I-Core program, the ISF, the BSF Transformative program, IMOS via ISA and by a Kimmel award. 

M.T.S. acknowledges support by a grant from IMOS/ISA, the Benoziyo center for Astrophysics at the Weizmann Institute of Science. This work was in part supported by the Scientific Discovery through Advanced Computing (SciDAC) program funded by U.S. Department of Energy Office of Advanced Scientific Computing Research and the Office of High Energy Physics. This research used resources of the National Energy Research Scientific Computing Center (NERSC), a U.S. Department of Energy Office of Science User Facility operated under Contract No. DE-AC02-05CH11231.

E.O.O is grateful for the support by grants from the Israel Science Foundation, Minerva, Israeli Ministry of Science, the US-Israel Binational Science Foundation, the Weizmann Institute and the I-CORE Program of the Planning and Budgeting Committee and the Israel Science Foundation.

C.~F gratefully acknowledges support of his research by the Heising-Simons Foundation (\#2018-0907).

A.A.M. is funded by the Large Synoptic Survey Telescope Corporation, the Brinson Foundation, and the Moore Foundation in support of the LSSTC Data Science Fellowship Program; he also receives support as a CIERA Fellow by the CIERA Postdoctoral Fellowship Program (Center for Interdisciplinary Exploration and Research in Astrophysics, Northwestern University)

MR has received funding from the European Research Council (ERC) under the European Union's Horizon 2020 research and innovation program (grant agreement n°759194 - USNAC).

\appendix
\section{Release of the {\tt PhotoManip} code}
The {\tt PhotoManip} tool, used to calculate the reference time for all the light curves and figures in this paper, is made available at {\tt https://github.com/maayane/PhotoManip}. 

The reference time is calculated as the epochs at which the extrapolated light curve is reaching zero. {\tt PhotoManip} fits either the $r$-band or the $g$-band flux during the rise time with an exponential function of the form
\begin{equation}\label{eq:bolometric luminosity exp}
f=f_{\rm{max}}\{ 1-\exp{[(t_{0}-t)/t_c]} \}\;,
\end{equation}
and a power-law of the form
\begin{equation}\label{eq:bolometric luminosity power}
f=a(t-t_{0})^n \;,
\end{equation}

(where $t_{0}$ is the time of zero flux, $f_{\rm{max}}$ is the maximum flux, $t_{c}$ is the characteristic rise time of the bolometric light curve). The fit uses the MCMC algorithm {\tt emcee}\citep{Foreman2013}.

\bibliographystyle{apj} 
\bibliography{bibliograph.bib}

%\LongTables
%\begin{landscape}

%

\end{document}